\documentclass[twocolumn,prb,showpacs]{revtex4}
\usepackage{graphicx}
\usepackage{dcolumn}
\usepackage{bm}

\begin{document}

\preprint{APS/123-QED}

\title{Effective Hamiltonian of Three-orbital Hubbard Model on Pyrochlore Lattice: Application to LiV$_2$O$_4$}

\author{K. Hattori}
\email{hattori@issp.u-tokyo.ac.jp}
\author{H. Tsunetsugu}%
\affiliation{%
The Institute for Solid State Physics, University of Tokyo, 5-1-5, Kashiwanoha, Kashiwa, Chiba 277-8581, Japan
}%

\date{\today}

\begin{abstract}
To investigate heavy fermion behavior in the 
vanadium spinel LiV$_2$O$_4$, we start from a three-orbital 
Hubbard model on the pyrochlore lattice and derive its 
low-energy effective Hamiltonian by an approach of 
real-space renormalization group type.  We first derive 
the effective Hamiltonian numerically, and then succeed
in representing the results into an analytic form with 
physical operators for low-energy degrees of freedom in 
tetrahedron unit.  The effective Hamiltonian is defined 
on the coarse grained lattice, i.e., face-centered cubic (f.c.c.)
 lattice , and it operates 
in a restricted Hilbert space defined in terms of 
a specific molecular orbital $T_2$ in the unit.  
One important tetrahedron configuration has a three-fold 
orbital degeneracy and spin $S=1$, and correspondingly, 
the effective Hamiltonian has spin and orbital 
exchange interactions of Kugel-Khomskii type as well as 
correlated electron hoppings.  
The coupling constants in the effective Hamiltonian 
are determined from the numerically obtained 
renormalized Hamiltonian, and also by means of perturbation.  
We calculate and analyze low-energy states 
of the effective Hamiltonian for 
the unit of four coupled tetrahedra both 
analytically and numerically.  
Effective hopping elements in the effective Hamiltonian 
are renormalized to about $1/10$ of the original hopping 
integral.  It is important that different virtual processes 
make opposite contributions to the exchange term, and consequently 
the coupling constant is given by a remaining small value.  
This is particularly prominent in the spin-spin channel, 
where ferromagnetic double exchange processes compete 
with antiferromagnetic superexchange processes.  
Another important point is that various spin and orbital 
exchange processes are competing to each other.  
Together with geometrical frustration of the effective 
f.c.c. lattice, these two features result in nearly degenerate 
three lowest-energy states of different types 
in the four coupled tetrahedra, and each of the three 
has a finite degeneracy in spin and/or orbital.  
We also calculate spatial correlations of spin and orbital 
and found that short-range spin-spin correlations are 
strongly entangled with orbital configurations.  
This indicates that a large remaining entropy at low 
temperature is related to slow coupled fluctuations of 
spin and orbital.  
These results suggest the absence of phase transition in spin 
and orbital spaces down to very low temperatures 
and their large fluctuations in the low-energy sector, 
which are key issues for understanding the heavy 
fermion behavior in LiV$_2$O$_4$.  
\end{abstract}

\pacs{71.27.+a, 71.10.Fd}  
\maketitle

\section{\label{sec:Intro}INTRODUCTION}
 The vanadium spinel LiV$_2$O$_4$ is the first heavy fermion compound discovered in d-electron systems \cite{1}. For about a decade, various experimental and theoretical efforts have been made to understand its heavy fermion behaviors. Low-temperature properties such as specific heat, magnetic susceptibility, electrical resistivity and Hall coefficient seem to be explained by the quasiparticle picture with a large effective mass.\cite{2,3,4,5} Corresponding to these low-temperature behaviors, the electronic spectral function develops a peak above the Fermi energy at low temperature observed in the laser photoemission spectroscopy \cite{6}. All these low-temperature behaviors are characterized by one energy scale $T^*\sim 30$ K. In contrast to these low-temperature properties, LiV$_2$O$_4$ exhibits bad metallic behaviors at higher temperatures\cite{7}. The temperature ($T$) dependence of magnetic properties is also interesting. The size of magnetic moment changes from a mixed-valent value (V$^{3+}$+V$^{4+}$) to a smaller value at about 500 K.\cite{2,8,9} Neutron experiment showed that spin fluctuations $\chi({\bf q})$ change their spatial correlations at around $T=T^*$. In the higher-temperature region, spin fluctuations show a broad peak at $\bf q=0$, and this is considered as a consequence of double exchange interactions. In the lower-temperature region, neutron experiments exhibited that the peak position shifts to $|{\bf q}|=|{\bf Q}^*|\sim 0.6 {\rm \ \AA}^{-1}$.\cite{10,11} Therefore, it is important to investigate the competition of ferro- and antiferro-magnetic fluctuations to understand the low-temperature heavy fermion behaviors.

Regarding theories, it has been discussed that the heavy fermion behaviors originate from the Kondo effect,\cite{12,13,14} the inter-orbital Coulomb interaction,\cite{15} the spin-orbital fluctuations,\cite{16,17} the frustrations,\cite{18,19} and dimensional crossover from coupled one-dimensional chains to three dimensions.\cite{20F}

 Band structure calculations show that the Fermi surfaces of LiV$_2$O$_4$ are composed by d-electrons $t_{\rm 2g}$ orbitals.\cite{12,12b,12c,24,12d} These $t_{\rm 2g}$ orbitals split into $a_{\rm 1g}$ and $e_{\rm g}$ orbitals due to the trigonal distortion of surrounding oxygen atoms. Anisimov { et al.} proposed the Kondo effect scenario\cite{12} that the $a_{\rm 1g}$ orbital plays a role of localized electron and interacts with conduction electrons in $e_{\rm g}$ orbitals. A recent study of cluster dynamical mean field theory\cite{21} claims that the heavy fermion behaviors are related to the criticality of orbital-selective Mott transition of $a_{\rm 1g}$ electrons based on the analysis using a simplified two-orbital Hamiltonian. Yushankhai { et al.} analyzed the low-temperature $\bf Q^*$ spin fluctuations observed in the neutron scattering experiment by employing a phenomenological self-consistent renormalization theory of spin fluctuation\cite{Yushankhai1,Yushankhai2}. They succeeded in fitting the neutron data qualitatively, but understanding of the microscopic aspects of magnetic fluctuations and heavy quasiparticles are desired. Despite of these efforts, the competition of ferro- and antiferro-magnetic interactions and crossover behaviors in the temperature dependence of susceptibility are not fully understood and it is desired to clarify how to describe the quasiparticles on a {\it frustrated} pyrochlore lattice and whether the frustration plays an important role for the realization of heavy fermion behaviors.

In this paper, we focus on the coupling of orbital degrees of freedom with spin and charge ones in LiV$_2$O$_4$\cite{16,17} starting with  a microscopic model on the pyrochlore lattice. We discuss its interplay with spin and charge degrees of freedom and its spatial correlations beyond a tetrahedron cluster. To examine the role of orbital degrees of freedom explicitly, we will use a three-orbital Hubbard model without assuming that $a_{1g}$ orbital is localized. Since the unit cell contains four vanadium atoms and each vanadium atom has three $t_{\rm 2g}$ orbitals, straightforward calculations are not applicable. In this paper, we shall employ a real space renormalization group approach to extract a low-energy effective Hamiltonian for tetrahedron units.\cite{TsuneHeisen} The effective Hamiltonian is $t$-$J$-like model: localized spin-one and orbital-triplet degrees of freedom  are coupled via exchange interactions and mobile electrons with three-fold orbital degeneracy hop between tetrahedron units. Using this effective model, we will discuss low-energy electron itineracy and competing interactions of spin and orbital degrees of freedom in LiV$_2$O$_4$.

This paper is organized as follows. In Sec. \ref{sec:Model}, we show the starting microscopic model used in this paper. Then in Sec. \ref{sec:Eigen1tet}, we will demonstrate the results of the exact diagonalization to find low-energy degrees of freedom in one tetrahedron unit. In Sec. \ref{sec:Effec1tet}, we will discuss a possible effective Hamiltonian which can describe the low-energy sector. In Sec. \ref{sec:4tet}, we will show the four-unit diagonalization results calculated by using the low-energy states in the one-tetrahedron calculations. In Sec. \ref{sec:exchangeTetra}, the low-energy physics is analyzed by the perturbative approach from the strong coupling limit. Finally, we discuss the effective model relevant to LiV$_2$O$_4$ in Sec. \ref{sec:Conclusion1} and summarize the present paper in Sec. \ref{sec:Conclusion2}.

\section{\label{sec:Model}MODEL}
\begin{figure}[t!]
  \begin{center}
    \includegraphics[width=.45\textwidth]{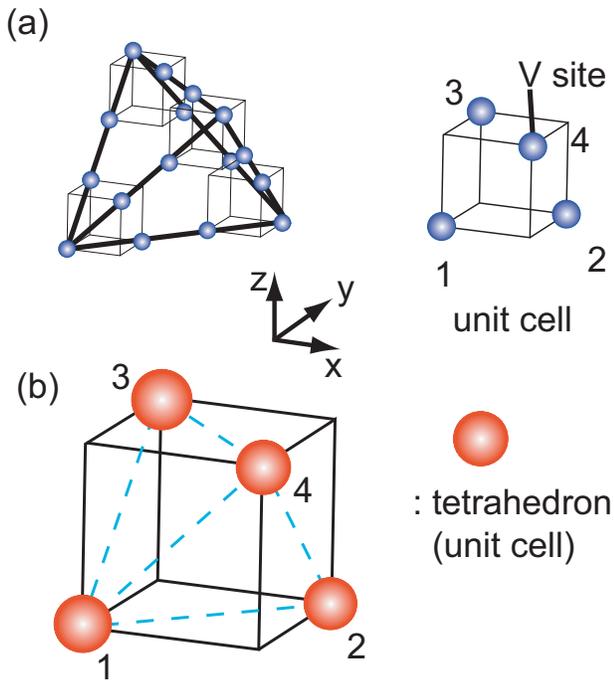}
  \end{center}
\caption{(Color online) (a) The lattice structure of pyrochlore lattice. The unit cell contains four sites forming a tetrahedron. (b) Tetrahedron units in the pyrochlore lattice form an face-centered cubic (f.c.c.) lattice.}
\label{fig-lat}
\end{figure}

We start with describing a realistic microscopic model of electronic structure for the vanadium spinel LiV$_2$O$_4$. In LiV$_2$O$_4$, the first principle band calculations\cite{12,12b,12c,24,12d} point out that the electronic density of states near the Fermi energy consists mainly of the d-electron $t_{2g}$ orbitals on vanadium sites. In the spinel structure, the vanadium sites form a three-dimensional pyrochlore lattice and the unit cell contains four vanadium atoms which form a tetrahedron as shown in Fig. \ref{fig-lat}. The electron hopping processes can be described by the effective V-V hoppings. Effects of V-O hoppings are included as a renormalization of V-V hoppings. There is trigonal distortion in the lattice due to O ion displacement. This lifts three-fold degenerate $t_{2g}$ orbitals into $a_{1g}$(singlet) and $e_g$(doublet). The vanadium valence is V$^{3.5+}$ in average and this corresponds to 1.5 electrons per atom, i.e., quarter filling of $t_{2g}$ orbital.

 The Hamiltonian we will investigate in this paper is a three-orbital $t_{\rm 2g}$ Hubbard model on the pyrochlore lattice with trigonal splittings,
\begin{eqnarray}
H\!\!\!\!&=&\!\!\!\!\sum_{{\bf i} {\bf j}}\sum_{\sigma\alpha\beta}t_{{\bf i}{\bf j}}^{\alpha\beta}d_{{\bf i}\alpha\sigma}^{\dagger}d_{{\bf j}\beta\sigma}+\sum_{{\bf i}\alpha\sigma}\Big[-\mu n_{{\bf i}\alpha\sigma}+\frac{U}{2}n_{{\bf i}\alpha\sigma}n_{{\bf i}\alpha-\sigma}\nonumber\\
&+&\!\!\!\! \sum_{\beta<\alpha}\sum_{\sigma'}\Big(U'n_{{\bf i}\alpha\sigma}n_{{\bf i}\beta\sigma'}+Jd^{\dagger}_{{\bf i}\alpha\sigma}d^{\dagger}_{{\bf i}\beta\sigma'}d_{{\bf i}\alpha\sigma'}d_{{\bf i}\beta\sigma}\Big)\Big],\label{Hamil}
\end{eqnarray}
where $d_{{\bf i}\alpha\sigma}^{\dagger}$ is a d-electron creation operator with the orbital $\alpha(=xy,yz$ or $zx)$ and the spin $\sigma(=\uparrow$ or $\downarrow$) at the site ${\bf i}$, and its number operator is defined as $n_{{\bf i}\alpha\sigma}=d_{{\bf i}\alpha\sigma}^{\dagger}d_{{\bf i}\alpha\sigma}$. The electron hoppings $t_{\bf ij}^{\alpha\beta}$ are limited to the nearest neighbor sites and $\mu$ is the chemical potential. The trigonal splittings are included in $t_{{\bf i}{\bf i}}^{\alpha\beta}$. For the interaction term, we use standard onsite Coulomb interactions without pair hopping terms as in other studies\cite{16,17}. The rotational symmetry of Coulomb interaction is satisfied by the relation $U=U'+J$ and we will use $U$, $U'$ and $J$ satisfying this condition throughout this paper. We choose the nearest neighbor tight-binding parameters $t_{{\bf ij}}^{\alpha\beta}$ by setting Slater-Koster parameters\cite{25} as $t_{\sigma}\equiv (dd\sigma)=-0.527\ {\rm eV}$, $t_{\pi}\equiv (dd\pi)=-0.085\sim -0.13\  {\rm eV}$, and $t_{\delta}\equiv (dd{\delta})=0.25 \ {\rm eV}$ for $\sigma$-, $\pi$-  and $\delta$-bond, respectively. We also fix the trigonal splitting $\Delta=\varepsilon_{e_{{\rm g}}}-\varepsilon_{a_{{\rm 1g}}}=0.02\ {\rm eV}$. Although $\Delta$ was estimated to be $\sim 0.1$ eV by the band calculation\cite{12}, it does not directly correspond to the ``microscopic" $\Delta$ we use.\cite{delta}

%

\begin{figure}[ht]
  \begin{center}
    \includegraphics[width=.5\textwidth]{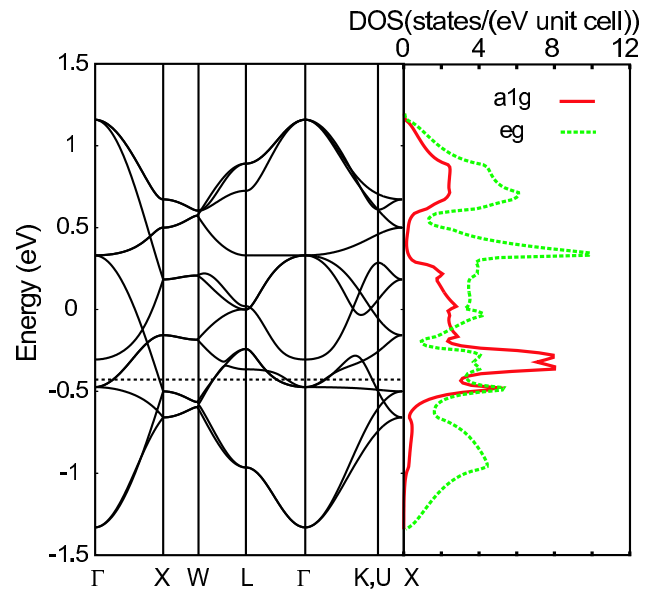}
  \end{center}
\caption{(Color online) Tight-binding dispersion and density of states calculated by the non-interacting Hamiltonian. The Fermi energy is indicated by dotted line.}
\label{fig-1}
\end{figure}
%

Before examining the effects of electron correlations, we study the electronic structure of the non-interacting case. Each unit cell contains four vanadium atoms and each atom has three orbitals and therefore there are twelve bands in total. Their energy dispersion and the non-interacting density of states are shown in  Fig. \ref{fig-1} for $t_{\pi}=-0.085$ eV. Apart from the high-energy regions the overall features are in good agreement with the first-principle band structure calculations\cite{12,12b,12c,24}. It is noted that the weight of $a_{1g}$ orbitals is larger than the one of $e_g$ near the Fermi energy. Among six electrons per unit cell, the occupation numbers are $n_{a_{1g}}=1.18$ and $n_{e_g}=4.82$ per tetrahedron in this parameter set. 

\begin{figure}[t]
  \begin{center}
    \includegraphics[width=.4\textwidth]{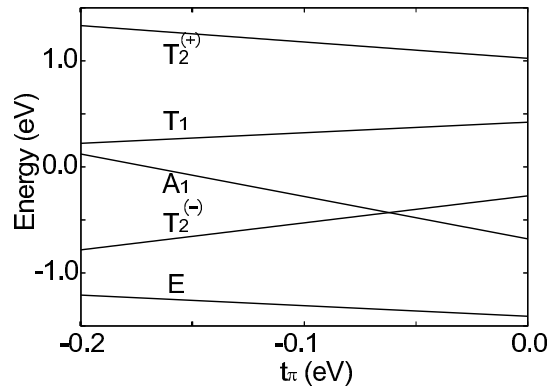}
  \end{center}
\caption{Non-interacting one-particle energies at $\Gamma$ point vs $t_{\pi}$.}
\label{fig-energyMol}
\end{figure}

For later purpose, let us first consider molecular orbitals in a single tetrahedron unit cell. Qualitatively, the energy levels of the molecular orbitals correspond to the band energies at the $\Gamma$ point. There are twelve molecular orbitals in total. The unit cell has the point group symmetry $T_d$. The twelve orbitals constitute five multiplets labeled by irreducible representations of $T_d$ group, $A_1$, $E$, $T_1$, $T_2^{(-)}$ and $T_2^{(+)}$. Since there are two $T_2$ representations, we distinguish them by $(-)$ and $(+)$. These orbitals are listed in Table \ref{tbl-1} and the wavefunctions of the molecular orbitals are shown in Appendix \ref{sec:1particleorbital}. We label the irreducible representations by $\Gamma$ and define the energy level as $\varepsilon_{\Gamma}$. In this paper, we choose $t_{\pi}$ value such that the $A_1$-level is higher than $T_{2}^{(-)}$-level and we show the $t_{\pi}$ dependence of the energy at the $\Gamma$ point in Fig. \ref{fig-energyMol}. We will see in Sec. \ref{sec:exchangeTetra} that the position of $A_1$- and $T_2^{(-)}$-levels are important for low-energy properties and this is sensitive to $t_{\pi}$. The parameter region we discuss in this paper is $t_{\pi}=-0.085 \sim -0.13$ eV and there $\varepsilon_{E}<\varepsilon_{T_2^{(-)}}<\varepsilon_{A_1}< \varepsilon_{T_1}< \varepsilon_{T_2^{(+)}}$. Large contribution of $a_{1g}$ orbital near the Fermi surfaces comes from $T_2^{(-)}$ and $A_1$ orbitals.

%

\begin{table}[t]
\caption{Molecular orbitals of tetrahedron unit and their energy. $D=[(\frac{3}{2}[t_{\sigma}+t_{\pi}]+t_{\delta}+\frac{\Delta}{6})^2+2(t_{\pi}-t_{\delta}-\frac{\Delta}{3})^2]^{\frac{1}{2}}$. The third and the fourth columns represent the weight of $a_{1g}$ and $e_g$ atomic orbitals, respectively for $t_{\pi}=-0.085$ eV. The values for two $T_2$ representations depend on the hopping and trigonal splitting parameters but their sums are constant.}
\begin{ruledtabular}

   \begin{tabular}{lccll}
         $\Gamma$& degeneracy & $\varepsilon_{\Gamma}$&$a_{\rm 1g}$& $e_{\rm g}$ \\
         \hline
       $T_2^{(+)}$ & 3 & $\frac{1}{4}t_{\pi}+\frac{3}{4}t_{\delta}-\frac{1}{12}\Delta+D$ & $0.35^*$ & $0.65^*$ \\
       $T_1$ & 3 & $-\frac{3}{4}t_{\sigma}+\frac{1}{2}t_{\pi}-\frac{3}{4}t_{\delta}+\frac{1}{6}\Delta$ &$0.00$   & $1.00$   \\
       $A_1$ & 1 & $\frac{3}{4}t_{\sigma}-2t_{\pi}+\frac{1}{4}t_{\delta}-\frac{2}{3}\Delta$ &$1.00$   & $0.00$   \\
       $T_2^{(-)}$ & 3 & $\frac{1}{4}t_{\pi}+\frac{3}{4}t_{\delta}-\frac{1}{12}\Delta-D$ & $0.65^*$ & $0.35^*$ \\
       $E  $ & 2 & $\frac{3}{4}t_{\sigma}-\frac{1}{2}t_{\pi}-\frac{5}{4}t_{\delta}+\frac{1}{6}\Delta$& $0.00$   & $1.00$   \\
   \end{tabular}
\end{ruledtabular}
\label{tbl-1}
\end{table}

\section{\label{sec:Eigen1tet}MANY-ELECTRON EIGENSTATES OF ONE TETRAHEDRON}

In this section, we include Coulomb interactions in the Hamiltonian (\ref{Hamil}) and investigate its eigenstates in a single tetrahedron. The results obtained in this section provide an insight to understand the high-temperature properties of LiV$_2$O$_4$. Moreover, the many-body wavefunctions obtained in this section become good bases for the low-energy effective model which will be discussed in Sec. \ref{sec:Effec1tet} and four-tetrahedron calculations in Sec. \ref{sec:4tet}. 

 In the first part of this section, we will show the energy spectra of one tetrahedron unit calculated by exact diagonalization for typical sets of parameters in the Hamiltonian (\ref{Hamil}). For LiV$_2$O$_4$, the average d-electron number is $1.5$ per vanadium site. This corresponds to six electrons per tetrahedron. We will discuss the energy spectra for the total d-electron number $n_d=4,5,6$ and 7. Then, in the second part, we will show the ground state phase diagram in the parameter space of Coulomb interactions. In the third part, we will also investigate how to construct low-energy spectra in terms of molecular orbital bases. In the final part of this section, we evaluate the temperature dependence of the thermodynamic quantities such as spin susceptibility, charge susceptibility and entropy.

\subsection{Energy spectra of one tetrahedron}\label{sec-EigenEnergy}
In this subsection, we show energy spectra of one tetrahedron unit. The low-energy eigenstates of one tetrahedron unit will be used as bases of the discussion in later sections. We will also discuss thermodynamic properties such as spin susceptibility and entropy, and in order to examine the thermodynamic quantities in the whole temperature region, we need all the important eigenstates. Since the Hilbert space is very large, we restrict ourselves in the subspace for total d-electron number $n_d\le 7$ in the tetrahedron unit and $d^0$, $d^1$, $d^2$, and $d^3$ configurations on each vanadium atom. The other configurations $d^n$ with $n\ge 4$ are not taken into account. We have checked the validity of this truncated calculation at least for the purpose of discussing the low-energy properties of this system by comparing low-energy eigenvalues calculated by this truncated calculation with those of the full Lanczos method. The configurations included in the truncated calculations contain physical processes such as super- and double-exchange interactions which are important when considering the low-energy properties of this system. The numerical diagonalizations were carried out with the open boundary condition, utilizing the spin rotational symmetry which reduces the maximum matrix size down to $\sim 46000$ for $n_d=7$ and the total spin $S=1/2$. 

\begin{figure}[!t]
  \begin{center}
    \includegraphics[width=.5\textwidth]{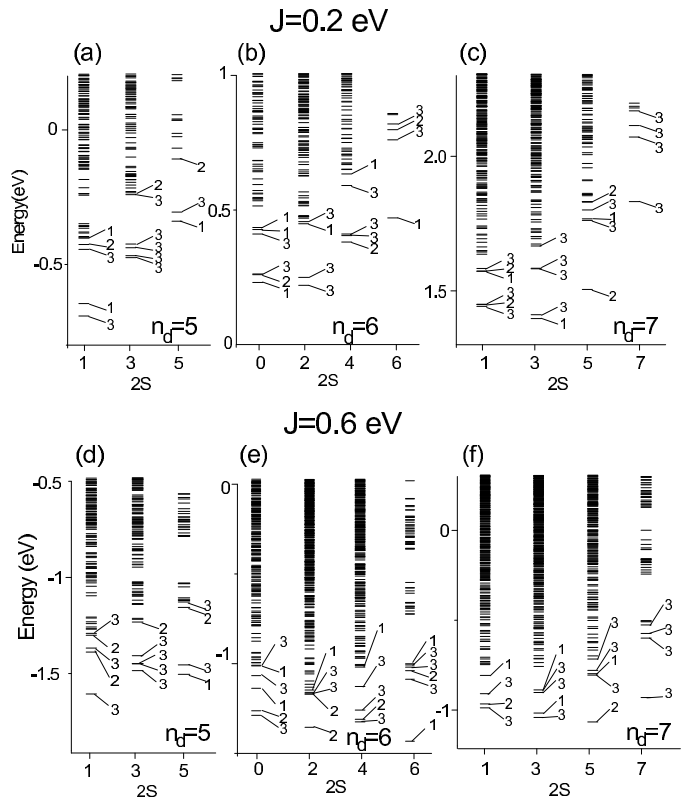}
  \end{center}
\caption{Energy spectra in each subspace of $n_d$ and spin ($2S$) for $U=1.5$ eV, $t_{\pi}=-0.085$ eV. (a) $n_d=5$, (b) $n_d=6$, (c) $n_d=7$, (d) $n_d=5$, (e) $n_d=6$, and (f) $n_d=7$. The numbers denote the orbital degeneracy. (a)-(c): $J=0.2$ eV, (d)-(e):$J=0.6$ eV.}
\label{fig-eneSpec}
\end{figure}

Figures \ref{fig-eneSpec} (a)-(c) show the energy eigenvalues in each subspace of total spin $S$ for $U=1.5$ eV, $U'=1.3$ eV, $J=0.2$ eV, and $t_{\pi}=-0.085$ eV. The numbers shown denote the degeneracy of each eigenstate that arises from the point group symmetry. The ground state of $n_d=6$ is total spin $S=1$ and orbital triplet. Figures \ref{fig-eneSpec} (d)-(f) show the energy eigenvalues for the same parameters in (a)-(c) except $J=0.6$ eV. As we increase $J$, the energies of the large spin states become lower. The ground state in $n_d=6$ sector is  fully polarized. In the $n_d=5$ and $7$ sectors, however, the ground states are not fully polarized states. It is noted that the energy differences between different $n_d$ for $J=0.6$ eV are smaller than those for $J=0.2$ eV. This implies that charge fluctuations are enhanced as $J$ is increased, which will be discussed in Sec. \ref{suscep}.

\subsection{Ground state phase diagram for one tetrahedron}\label{GS1tet}
\begin{figure}[!t]
  \begin{center}
    \includegraphics[width=.5\textwidth]{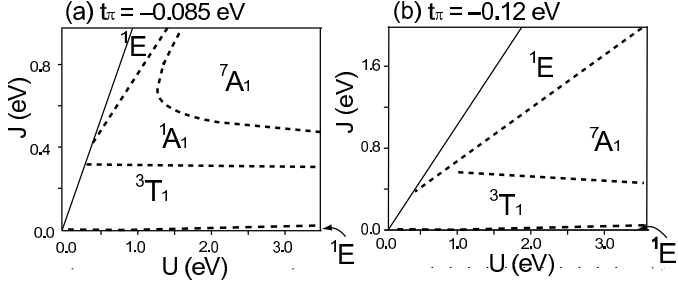}
  \end{center}
\caption{Ground state phase diagram for one tetrahedron with $n_d=6$. $U'=U-J$. (a) $t_{\pi}=-0.085$ eV, and (b) $-0.12$ eV. The dotted lines are the boundary between different ground states. The large-$J$ region in (a) ($J>1.0$ eV) is similar to the corresponding region in (b)}
\label{fig-3}
\end{figure}

\begin{figure}[t]
  \begin{center}
    \includegraphics[width=.3\textwidth]{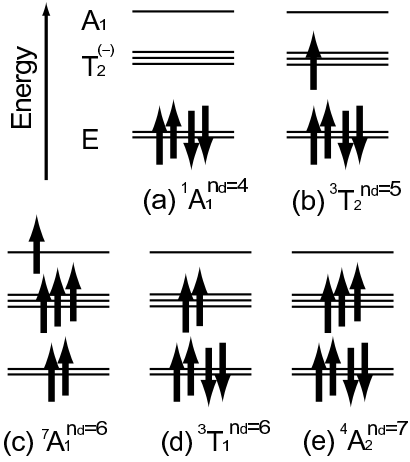}
  \end{center}
\caption{Schematic interpretation of ground states in the molecular orbital picture.}
\label{fig-schem}
\end{figure}
%

\begin{figure}[!t]
  \begin{center}
    \includegraphics[width=.5\textwidth]{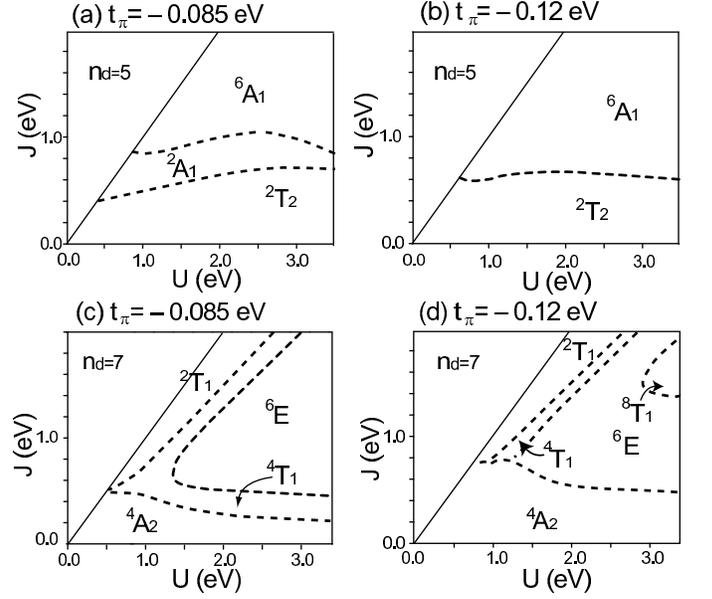}
  \end{center}
\caption{Ground state phase diagram for $n_d=5$ and $7$. $U'=U-J$. (a) $n_d=5$ and $t_{\pi}=-0.085$ eV. (b) $n_d=5$ and $-0.12$ eV. (c) $n_d=7$ and $t_{\pi}=-0.085$ eV. (d) $n_d=7$ and $t_{\pi}=-0.12$ eV. The dotted lines are the boundary between different ground states.}
\label{fig-3n=57}
\end{figure}
%
 In Fig. \ref{fig-3}, we show the ground state phase diagram of the $n_d=6$ space in the $U-J$ plane for $t_{\pi}=-0.085$ and $-0.12$ eV. Note that the region $J>U$ is unphysical. These results are obtained for the full Hamiltonian (\ref{Hamil}) without truncating the Hilbert space. We use a usual Lanczos method to calculate the eigenenergies of the ground and the first excited states. 
There are five phases dependent on $U$, $J$ and $t_{\pi}$. Their total spin $S$ and point group irreducible representation $\Gamma$ are determined. We represent the eigenstates by a usual notation $^{2S+1}\Gamma$ and if necessary, we will also write the electron number explicitly as $^{2S+1}\Gamma^{n_d}$. The five phases correspond to $^1\!E$, $^7\!A_1$, $^1\!A_1$, $^3T_1$ and another $^1\!E$. On increasing $|t_{\pi}|$, $^1\!A_1$ state disappears. The other states seem to be robust against the variation of $t_{\pi}$. 

A wide range of the phase space is covered by $^7\!A_1$ and $^3T_1$ states. The $^7\!A_1$ state has a fully-polarized spin moment $S=3$ which arises from ferromagnetic double-exchange interactions. The $^3T_1$ state is stabilized in the competition between double-exchange and antiferro-magnetic super-exchange interactions. Electron configurations of these ground states are schematically depicted in  Fig. \ref{fig-schem}. As is easily seen in (c), the $^7\!A_1$ phase corresponds to the case of ``strong'' Hund's coupling; $J$ is larger than the level separations. As shown in (d), the $^3T_1$ phase corresponds to the case of ``moderate'' Hund's coupling, since the Hund's coupling is effective only in the $T_2^{(-)}$ molecular orbitals. There appear two $^1\!E$ states in the phase diagram. These states are located at the region where $J$ is too small or large so that it would not be important to discuss the properties of LiV$_2$O$_4$. Therefore, we do not consider these $^1\!E$ states in more detail in the following.

 To compare with the $n_d=6$ case, we also show the ground state phase diagram for $n_d=5$ and $7$, in Fig. \ref{fig-3n=57}. As is seen, $^6\!A_1^{n_d=5}$ and $^2T_2^{n_d=5}$ states, and $^6E^{n_d=7}$ and $^4\!A_2^{n_d=7}$ states are the ground states in a wide range of the parameters. It is illuminating to notice that the $^4\!A_2^{n_d=7}$ state is obtained by adding an electron to the $T_2^{(-)}$ molecular orbital in the $^3T_1^{n_d=6}$ state and the $^2T_2^{n_d=5}$ state is obtained by removing an electron from the $T_2^{(-)}$ orbital. This is easily understood in Fig. \ref{fig-schem}. In a similar way, $^6E^{n_d=7}$ state is obtained by adding an electron to the $E$ molecular orbital in the $^7\! A_1^{n_d=6}$ state and $^6\!A_1^{n_d=5}$ state is obtained by removing an electron from the $A_1$ orbital. 
The fully polarized state $^8T_1^{n_d=7}$ does not appear in Fig. \ref{fig-3n=57} (c). With increasing $|t_{\pi}|$, the energy of $^8T_1^{n_d=7}$ state decreases around the area $U > 3.0$ eV and $J\sim1.7$ eV and then becomes the ground state as shown in  Fig. \ref{fig-3n=57} (d). $^2T_1^{n_d=7}$ state corresponds to the state obtained by adding a $T_2^{(-)}$ molecular orbital electron to $^1\!E^{n_d=6}$. As in the cases of $^1\! E^{n_d=6}$, we will not further discuss the region where $J$ is large.

\subsection{Low-energy spectrum of one tetrahedron}\label{sec:shell}
\begin{figure}[!t]
  \begin{center}
    \includegraphics[width=.4\textwidth]{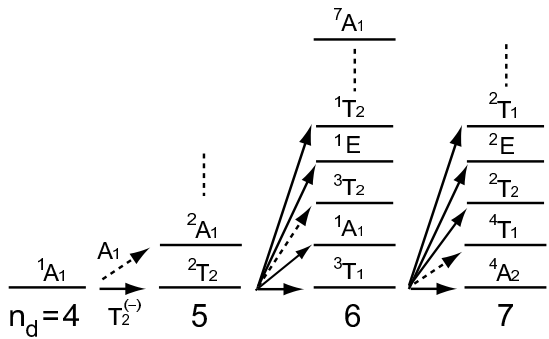}
  \end{center}
\caption{Low-energy level scheme and the quantum numbers $^{2S+1}\Gamma$ in each $n_d$.  $U=1.5$ eV, $J=0.2$ eV and $t_{\pi}=-0.085$ eV. Arrows shown by solid (dotted) line represent $T_2^{(-)}$ ($A_1$) one-particle excitations. For simplicity, we show only the excitations from the ground states for each $n_d$.}
\label{fig-config}
\end{figure}
%

In this subsection, we proceed to detailed investigation of the electron configurations of low-energy states. We will justify the schematic picture of Fig. \ref{fig-schem} by checking whether the same picture applies to the ground states in other $n_d$ subspaces. We will concentrate on the $^3T_1^{n_d=6}$ ground states since as will be shown in Sec. \ref{suscep}, the high-temperature thermodynamic properties of this phase are similar to the experimental data of LiV$_2$O$_4$.

The low-energy part of Figs. \ref{fig-eneSpec} (a)-(c) is schematically shown in Fig. \ref{fig-config}. Not shown in Fig. \ref{fig-eneSpec}, the ground state of $n_d=4$ is $^1\!A_1$ and the energy gap to the first excited states is large, $0.253$ eV. Considering this and quantum numbers, the ground state of $n_d=4$ can be considered as a ``closed shell'' state which corresponds to the fully occupied $E$ orbitals in the sense of the schematic picture in Fig. \ref{fig-schem} (a). Starting from this, the low-energy spectra for $n_d\ge 5$ can be successively constructed by adding electrons in the $T_2^{(-)}$ or $A_1$ molecular orbital as depicted in Fig. \ref{fig-config}. Note that all the ground states are constructed by adding electrons in the $T_2^{(-)}$ orbital. Indeed, the following group theoretical arguments justify this picture. 

In order to characterize the low-energy spectrum in more detail, it is important to identify the quantum numbers of the ground states in each $n_d$ subspace shown in Fig. \ref{fig-config}. As for the total spin, it increases by $1/2$ upon adding one electron. This simply means that electrons in the $T_2^{(-)}$ orbitals tend to align their spins, i.e., the Hund's rule. The symmetry of the orbital part can be also understood by starting from ground state in the $n_d=4$ subspace. This has the closed-shell electron configuration and therefore the symmetry of $A_1$-representation. The ground state in the $n_d=5$ subspace is constructed by adding one electron in the $T_2^{(-)}$ orbitals. Its symmetry is given by the product of two representations, one for the starting many-body wavefunction and the other for the molecular orbital of added electron. This case is simple and the result is $A_1\otimes T_2=T_2$. 

Next, the ground state for $n_d=6$ is constructed similarly by adding the second electron in the $T_2^{(-)}$ orbitals to the ground state $T_2^{n_d=5}$. Thus, considering the decomposition of product representation  $T_2\otimes T_2=A_1\oplus E\oplus T_1\oplus T_2$, we have four possibilities of orbital symmetry for the ground state in $n_d=6$ space. However, the spin part is triplet ($S=1$) due to the Hund's rule and its wavefunction is symmetric, and therefore we should choose an antisymmetric representation in $T_2\otimes T_2$. This is indeed unique and $T_1$. Thus, the wavefunction of $n_d=6$ with $S=1$ should have a $T_1$ symmetry if the $T_2^{(-)}$ orbital plays a role of the one-particle excitations. The other states, $^1\!A_1,\ ^1\!E$ and $^1T_2$ appear as excited states as shown in Fig. \ref{fig-config}. 

Finally, the ground state for $n_d=7$ is, once again, constructed by adding the third electron in the $T_2^{(-)}$ orbitals. Since the ground-state wavefunction has spin $3/2$ as predicted by the Hund's rule, all the three $T_2^{(-)}$ orbitals are occupied by electrons, aligning their spins. The symmetry of this state is $^4\!A_2$. In the group theoretical language, this $A_2$ is understood from the relation $T_1\otimes T_2 = A_2\oplus E\oplus T_1\oplus T_2$. The states other than $A_2$ also appear as excited states as shown in Fig. \ref{fig-config}.

\subsection{Spin susceptibility and entropy}\label{suscep}
Experiments of magnetic susceptibility indicate that, at around $T\sim 500$ K, the effective moment of vanadium ion $S_{\rm eff}^2$ changes from $\sim 1.5$-$1.75$ at high temperatures to $\sim$0.9 at low temperatures\cite{8,9}. This behavior was interpreted as Kondo like screening by Hopkinson {et al.,}\cite{14} but it is important to check whether alternative explanations are possible. The Weiss temperature $\Theta$ also changes at $\sim 500$ K\cite{8,9}; it is estimated as $\Theta\sim -500$ K by fitting the results at $600$-$1000$ K and $\Theta'\sim -30$ K at $80$-$300$ K. It is valuable to calculate the temperature dependence of susceptibility  from our one-tetrahedron data and compare the results to the experimental data.

The spin susceptibility $\chi_s(T)$ per vanadium site is given by $\chi_s(T)=g^2\mu_B^2\langle S^2\rangle/3N_sT$ with $g$, $\mu_B$, $N_s$, the electron's g-factor, Bohr magneton and the number of sites, respectively. Here, $\langle\cdots\rangle$ denotes grand canonical average at temperature $T$ with keeping the average electron number at 1.5 per site, and we evaluate this by averaging over the truncated Hilbert space of $n_d=5$, $6$ and $7$ as explained in Sec. {\ref{sec-EigenEnergy}}. In order to obtain $\chi_s(T)$, we calculate $\langle S^2\rangle$ with varying temperature.

\begin{figure}[t]
  \begin{center}
    \includegraphics[width=.45\textwidth]{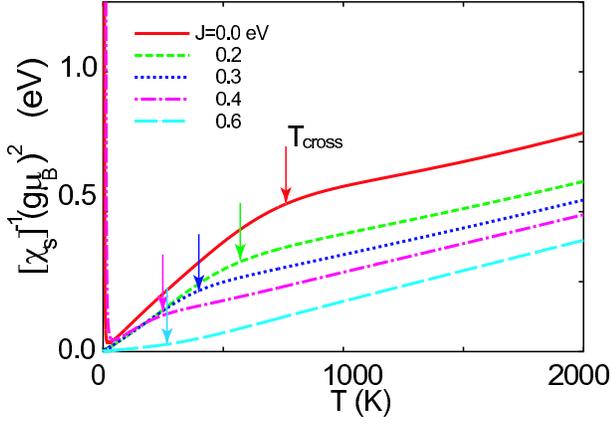}
  \end{center}
\caption{(Color online) Temperature dependence of inverse spin susceptibility.  $U=1.5$ eV, $t_{\pi}=-0.085$ eV and $U'=U-J$. The arrows indicate cross-over temperatures $T_{\rm cross}$.}
\label{fig-2}
\end{figure}

 Figure \ref{fig-2} shows the temperature dependence of the inverse spin susceptibility $\chi_s^{-1}(T)$ for several $J$ values. In the high-temperature regime, the $\chi_s(T)$ follows a Curie-Weiss law irrespective of $J$ values. There is a clear crossover marked by arrow at the temperature $T_{\rm cross}$, where the slope of $\chi_s^{-1}(T)$ changes. The high-temperature Weiss temperatures $\Theta$ is estimated in the region of $T\ge 800$ K  and the low-temperature Weiss temperature $\Theta'$ is estimated in the region of 50-200 K. The results are listed in Table \ref{tbl-2} together with the magnitude of the effective moment per site $S^2_{\rm eff}$ ($S'^2_{\rm eff}$). $\Theta$ ($\Theta'$) and $S^2_{\rm eff}$ ($S'^2_{\rm eff}$) are estimated by using the following form:
\begin{eqnarray}
\chi_s(T)=\frac{g^2\mu_B^2S_{\rm eff}^2}{3(T-\Theta)}.
\end{eqnarray}

\begin{table}[b]
\caption{Curie-Weiss temperature and effective moments. $\Theta$ and $S^2_{\rm eff} $ are estimated by fitting data in Fig. \ref{fig-2} at $T\ge 800$ K. Except $J=0.4$ eV, $\Theta'$ and $S'^2_{\rm eff} $ are estimated by fitting data in Fig. \ref{fig-2} at $50$-$200$ K. $\Theta'$ and $S'^2_{\rm eff} $ for $J=0.4$ eV are estimated at $90$-$160$ K. }
\begin{ruledtabular}
   \begin{tabular}{cccccc}
       $J$ (eV) & $0.0$ & $0.2$& $0.3$&$0.4$ & $0.6$ \\
         \hline
      $\Theta$ (K) & $-2133$ & $-1233$& $-774$&$-395$ & $171$ \\
      $S^2_{\rm eff} $ (V$^{-1}$) & $1.43$ & $1.43$& $1.39$&$1.33$ & $1.23$ \\
      $\Theta'$ (K) & $-4.5$ & $6.1$& $2.2$&$-27$ & $2.5$ \\
      $S'^2_{\rm eff} $ (V$^{-1}$) & $0.327$ & $0.450$& $0.456$&$0.529$ & $2.91$
   \end{tabular}
\end{ruledtabular}

\label{tbl-2}
\end{table}
The magnitude of high-temperature $S^2_{\rm eff}$ is comparable to $(3/4+2)/2=1.375$ which corresponds to that of the atomic mixed-valence limit with the Hund's rule, $d^1$ with the spin $s=1/2$ and $d^2$ with $s=1$. It is noted that the high-temperature $S^2_{\rm eff}$ has only weak $J$ dependence. $\chi_s(T)$ for $J=0.3$ eV is qualitatively in good agreement with the experimental results\cite{8,9} in three points: $\Theta$, $T_{\rm cross}$ and $S^2_{\rm eff}$. The experimental values are $\Theta\sim -500$ K, $T_{\rm cross}\sim 500$ K and $S^2_{\rm eff}\sim 1.5$-$1.75$\cite{8,9}. The deviations of $S_{\rm eff}^2$ would be due to the ferromagnetic correlations beyond the present calculations. $|\Theta'|$ and $S'^2_{\rm eff}$ are smaller than high-temperature $|\Theta|$ and $S^2_{\rm eff}$, respectively. The values of $S'^2_{\rm eff}$ reflect the ground and low-lying excited states. For $J=0.2$, 0.3 and 0.6 eV, $S'^2_{\rm eff}$ is very close to that for the ground states. Although the ground states are spin singlet for $J=0.0$ and 0.4 eV, the low-energy excited states with spin $S\ne 0$ contribute $S^2_{\rm eff}> 0$ in the temperature range where we fit the data. The $\chi_s(T)$ for $J=0.6$ eV has a clear ferromagnetic behavior due to the double exchange mechanism, see $\Theta > 0$ in Table \ref{tbl-2}. The behaviors at very low temperatures depend on the total spin $S$ of the ground state for each parameter set. The low-temperature upturns for $J=0.0$ and $0.4$ eV reflect that the ground states are spin-singlet states. 
\begin{figure}[t]
  \begin{center}
    \includegraphics[width=.45\textwidth]{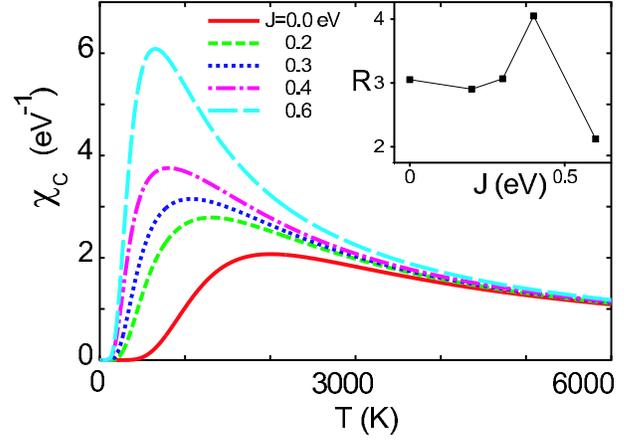}
  \end{center}
\caption{(Color online) Temperature dependence of charge susceptibility. $U=1.5$ eV, $t_{\pi}=-0.085$ eV and $U'=U-J$. Inset: $J$ dependence of $R=T^{\rm max}/T_{\rm cross}$.}
\label{fig-2chiC}
\end{figure}
%
\begin{figure}[t]
  \begin{center}
    \includegraphics[width=.45\textwidth]{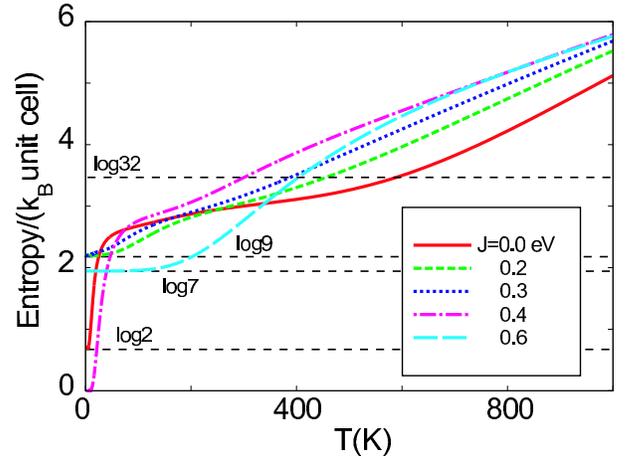}
  \end{center}
\caption{(Color online) Entropy per tetrahedron unit vs temperature. $U=1.5$ eV, $t_{\pi}=-0.085$ eV and $U'=U-J$.}
\label{fig-ent}
\end{figure}
%

We find that the cross-over temperature marked by arrow in Fig. \ref{fig-2} is related to the energy scale of charge fluctuations. The charge susceptibility $\chi_c(T)\equiv\langle n_d^2 -\langle n_d\rangle^2\rangle/T$ is shown in Fig. \ref{fig-2chiC}. The peak position of $\chi_c(T)$ $(\equiv T^{\rm max})$ is related to $T_{\rm cross}$ of spin susceptibility in Fig. \ref{fig-2}. The inset of Fig. \ref{fig-2chiC} shows the ratio $R\equiv T^{\rm max}/T_{\rm cross}$. Apart from large $J$ region, $R$ is nearly constant and $R \sim 3$. From this, we can understand that the crossover in spin susceptibility arises from charge fluctuations at least from small to moderate $J$ values. This interpretation is consistent with the effective moments in Table \ref{tbl-2} and suggested by the early exact diagonalization study\cite{23}. The interpretation of the crossover attributed to charge fluctuations in a tetrahedron is valid at least when tetrahedron coupling is weak, e.g., at high temperature. 

Figure \ref{fig-ent} shows the entropy ${\mathcal S}(T)$ per tetrahedron as a function of temperature. The finite values at zero-temperature are due to the degeneracy of ground states in $n_d=6$ subspace. The experimental data of entropy at $100$ K from the specific heat data is $\sim 5k_B\log2$ ($k_B$: Boltzmann constant) per tetrahedron,\cite{3} and this is larger than the present results at 100 K. Since we ignore inter-tetrahedron correlations in the present calculations, it is not adequate to discuss the low-temperature entropy quantitatively. We note that there still remains large entropy (more than $k_B\log 9$ at 100 K) for $J\le 0.4$ eV. This low-energy entropy might become an origin for heavy fermion behaviors in LiV$_2$O$_4$. This point will be discussed in Sec. \ref{sec:1tetHeff} based on an effective model for coupled tetrahedra.

\section{\label{sec:Effec1tet}LOW-ENERGY EFFECTIVE MODEL OF ONE TETRAHEDRON UNIT}
 In order to discuss low-energy properties of LiV$_2$O$_4$, one has to notice that among several ground states of one tetrahedron unit, $^3T_1$ phase has both spin and orbital degrees of freedom. The magnetic susceptibility and the entropy calculated for this phase capture the character of the experimental results at high temperature. Therefore, we now focus on the $^3T_1$ phase and discuss its low-energy properties in detail. To describe metallic behaviors of LiV$_2$O$_4$, it is important to examine one-particle excitations in this phase. We will construct an effective Hamiltonian for one tetrahedron unit and demonstrate that the $T_2^{(-)}$-orbital electrons only are sufficient to describe the low-energy one-particle excitations of Hamiltonian (\ref{Hamil}) in $^3T_1$ phase. This construction can be regarded as a procedure of a real-space renormalization group.\cite{TsuneHeisen} Based on the results obtained in this section, we will proceed to the next procedure of  the renormalization group in Sec. \ref{sec:4tet}.
\begin{figure}[t]
  \begin{center}
    \includegraphics[width=.5\textwidth]{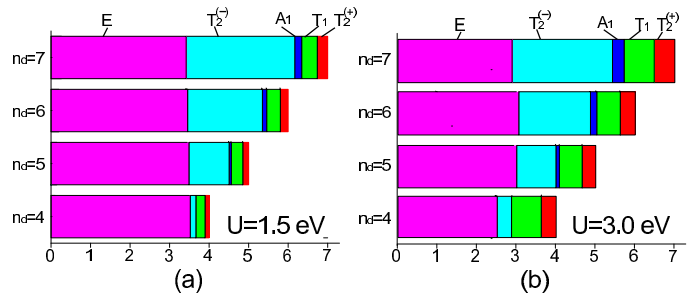}
  \end{center}
\caption{(Color online) Occupation numbers of molecular orbitals in the ground states in each $n_d$ for $J=0.3$ eV and $t_{\pi}=-0.12$ eV. (a) $U=1.5$ eV. (b) $U=3.0$ eV.}
\label{fig-molN}
\end{figure}
%

\subsection{One-particle excitations}\label{sec:1particle}
 Let us now investigate one-particle excitations in the $^3T_1$ phase in detail. First, we examine which molecular orbitals in a tetrahedron play a dominant role in the one-particle excitations upon changing electron number $n_d=6 \to 7$ and $n_d=6\to 5$. To this end, we define matrix elements $A_{\Gamma\Gamma'}^{n_d}$ by
\begin{equation}
A_{\Gamma\Gamma'}^{n_d}\equiv\sum_{g_{n_d}g_{n_d+1}}\langle g_{n_d+1}| d^{\dagger}_{\Gamma\uparrow}|g_{n_d}\rangle\langle g_{n_d}| d_{\Gamma'\uparrow}|g_{n_d+1}\rangle. \label{A}
\end{equation}
Here, $d_{\Gamma\uparrow}^{\dagger}$ with $\Gamma=E,T_2^{(-)},A_1,T_1$ and $T_2^{(+)}$ represents the d-electron creation operator in the molecular orbital basis with the spin $\sigma=\uparrow$. $g_{n_d}$ denotes the ground states in the $n_d$ subspace. $A_{\Gamma\Gamma'}^{n_d}$ can be easily calculated for the noninteracting case. We calculated $A_{\Gamma\Gamma'}^5$ and $A_{\Gamma\Gamma'}^6$, for the $^3T_1$ phase at $U=1.5$ eV, $J=0.2$ eV and $t_{\pi}=-0.12$ eV. The matrix element $\langle g_{n_d+1}| d^{\dagger}_{\Gamma\sigma}|g_{n_d}\rangle$ for $\Gamma= T_2^{(-)}$ is about ten times larger in magnitude than the others. We find that the largest eigenvalue $\lambda_{\rm max}$ of $A_{\Gamma\Gamma'}^{n_d}$ is suppressed to about 80 \% of the noninteracting value $\lambda_{\rm max}^{\rm free}$
\begin{eqnarray}
&&(n_d=5)\ \ \ \lambda_{\rm max}\sim 1.6<\lambda_{\rm max}^{\rm free}=2,\\
&&(n_d=6)\ \ \ \lambda_{\rm max}\sim 0.80<\lambda_{\rm max}^{\rm free}=1.
\end{eqnarray}
It is important that the eigenvector of the largest eigenvalue has almost all weights in $T_2^{(-)}$ components ($\sim 99$\%). Thus, we obtain the short-range contribution of renormalization factor $Z\sim 0.8$ in the present one tetrahedron calculation. The result for $U=3.0$ eV, $J=0.2$ eV and $t_{\pi}=-0.12$ eV is that the matrix element $|\langle g_{n_d+1}| d^{\dagger}_{\Gamma\sigma}|g_{n_d}\rangle|$ for $\Gamma= T_2^{(-)}$ is still about ten times larger than others and we obtain $Z\sim 0.66$. These results indicate that the $T_2^{(-)}$ orbitals play a dominant and important role for one-particle excitations between the ground states of each $n_d$ subspace.

We also calculate the electron occupation number of each orbital $\sum_{\alpha\in\Gamma}\sum_{\sigma}\langle  g_{n_d}|d_{\alpha\sigma}^{\dagger}d_{\alpha\sigma}|g_{n_d}\rangle$ for each $n_d$ subspace, and the result is plotted in Fig. \ref{fig-molN}. The value is averaged over degenerate ground states. The occupation number of the $T_2^{(-)}$ orbital increases by nearly one when $n_d$ increases by one. This is consistent with the analysis of $A_{\Gamma\Gamma'}^{n_d}$. As discussed in Sec. \ref{GS1tet}, a simple picture of the $^3T_1^{n_d=6}$ ground state is the fully occupied $E$ orbitals plus partially filled $T_2^{(-)}$ orbitals. The result of orbital occupation confirms this picture but also shows that a non-negligible  number of electrons occupy the high-energy one-particle molecular orbitals such as $T_1$ and $T_2^{(+)}$, suppressing the $E$-orbital occupation from four. The larger correlations $U$, the larger number of electrons occupy the high-energy molecular orbitals as shown in Fig. \ref{fig-molN} (b). Thus, the one-particle excitations should be regarded as dressed quasiparticles in the Landau's Fermi liquid picture rather than ``free'' electrons.

\begin{figure}[t]
  \begin{center}
    \includegraphics[width=.5\textwidth]{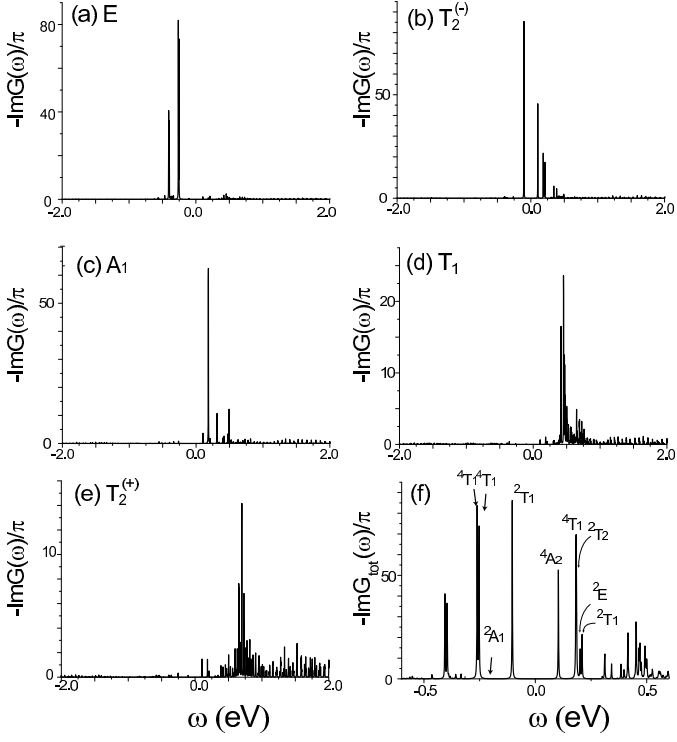}
  \end{center}
\caption{One-particle spectrum $-{\rm Im}G_{\Gamma}(\omega+i\delta)$ vs $\omega$. $U=1.5$ eV, $J=0.3$ eV, $t_{\pi}=-0.12$ eV and $\delta=0.001$ eV. (a) $\Gamma= E$, (b) $T_2^{(-)}$, (c) $A_1$, (d) $T_1$, and (e) $T_2^{(+)}$. (f) Detail structure of the low-energy region. The quantum number $^{2S+1}\Gamma$ for the corresponding peak is also shown. }
\label{fig-Gf}
\end{figure}
%

These investigations are directly checked by calculating Green's functions $G_{\Gamma}(\omega+i\delta)$ in the molecular-orbital basis, where $\delta$ is infinitesimal constant. $G_{\Gamma}(\omega+i\delta)$ is defined as a Fourier transform of a retarded Greens function $G_{\Gamma}(t)$ at $T=0$ 
\begin{eqnarray}
G_{\Gamma}(t)=-i\theta(t)\sum_{g_{n_d}}\langle g_{n_d}|\{d_{\Gamma}(t),d_{\Gamma}^{\dagger}(0)\}|g_{n_d}\rangle\ \ \ (n_d=6),
\end{eqnarray}
where $\theta(t)$ is Heaviside's step function, $\{\cdots\}$ denotes anticommutator and we omit the spin index $\sigma$.
We show the one-particle spectral function $-{\rm Im}G_{\Gamma}(\omega+i\delta)/\pi$ with $\delta=0.001$ eV in Figs. \ref{fig-Gf} (a)-(e). Note that the scale of the vertical axis is different for each figure. There are large peaks in low-energy region for $\Gamma=E ,\ T_2^{(-)}$  and $A_1$. On the other hand, there are no large peak in low-energy region for $\Gamma=T_1$ and $T_2^{(+)}$ but broad incoherent components in the high-energy region. This also agrees with the simple picture in Fig. \ref{fig-schem}. In Fig. \ref{fig-Gf} (f), we show the low-energy part of total spectral weight $-{\rm Im}G_{\rm tot}(\omega+i\delta)/\pi$ defined as $G_{\rm tot}(\omega+i\delta)\equiv\sum_{\Gamma}G_{\Gamma}(\omega+i\delta)$. As expected from Fig. \ref{fig-config}, the lowest-energy excitations are those of $T_2^{(-)}$ orbital and the corresponding peaks are very large. Another important point is that the peak of the $A_1$ orbital is also large and located at low energy.

\subsection{Effective Hamiltonian of one tetrahedron unit}\label{sec:1tetHeff}
The results obtained in the previous subsection show that $T_2^{(-)}$ orbitals only are sufficient to describe the low-energy sector of $^3T_1$ phase. In this subsection, we construct an effective Hamiltonian of these $T_2^{(-)}$ electrons, and determine its interaction parameters.

The general Hamiltonian with spin rotation symmetry of $T_2^{(-)}$ electrons at the tetrahedron $n$, retaining only the two-body interactions, should be written as
\begin{eqnarray}
H_{\rm eff}^{\rm 1tet}(n)\!\!&=&\epsilon\sum_{\sigma\alpha}n_{n\alpha\sigma}+\tilde{U}\sum_{\alpha}n_{n\alpha\uparrow}n_{n\alpha\downarrow}\nonumber\\
&+&\sum_{\alpha > \beta}\sum_{\sigma\sigma'}
\Big(\tilde{U}'n_{n\alpha\sigma}n_{n\beta\sigma'}+\tilde{J}a^{\dagger}_{n\alpha\sigma}a^{\dagger}_{n\beta\sigma'}a_{n\alpha\sigma'}a_{n\beta\sigma}\Big)\nonumber\\
&+&\tilde{T}\sum_{\alpha\ne\beta}a^{\dagger}_{n\alpha\uparrow}a^{\dagger}_{n\alpha\downarrow}a_{n\beta\downarrow}a_{n\beta\uparrow}+C,\label{Heff}
\end{eqnarray}
where $C$ is a constant and $n_{n\alpha\sigma}=a_{n\alpha\sigma}^{\dagger}a_{n\alpha\sigma}$ with $\alpha=a,b$ or $c$ (see, Appendix \ref{sec:1particleorbital}). $a^{\dagger}_{n\alpha\sigma}$ creates a ``quasiparticle'' of $T_2^{(-)}$ orbital at a tetrahedron $n$ that is dressed by the interactions and its vacuum corresponds to the $^1\!A_1^{n_d=4}$. $\epsilon$ is the one-particle energy level. The interaction parameters $\tilde{U},\ \tilde{U}'$ and  $\tilde{J}$ are the molecular-orbital version of the coupling in the $t_{2g}$ Hubbard model (\ref{Hamil}) and now the pair hopping term $\tilde{T}$ is also generally generated.

In order to check the validity of Hamiltonian (\ref{Heff}), we compare the numerically calculated eigenenergies $(E_{\rm num})$ of four-site case of the original Hamiltonian (\ref{Hamil}) with one-``site'' eigenenergies of the effective model (\ref{Heff}).  The low-energy eigenvalues are listed in Table \ref{tbl-4}. The number in the fifth and sixth columns is $\langle H_{\rm eff}^{{\rm 1tet}}\rangle/E_{{\rm num}}$, which measures the validity of Hamiltonian (\ref{Heff}). The results are  very close to unity and the validity of Hamiltonian (\ref{Heff}) is quantitatively proved. The estimated values of the interaction parameters turn out to be smaller than the bare d-electron interactions by the factor $1/10\sim1/5$. This is because the orbitals are extended over four sites and there is a reduction of energy scale by the one-tetrahedron renormalization factor $Z$ as discussed in Sec. \ref{sec:1particle}. The pair hopping term $\tilde{T}$ is also induced in this effective model but its strength is weaker than the others.

We note that a few states in Table \ref{tbl-4} cannot be described by only $T_2^{(-)}$ orbitals, and these states correspond to $A_1$-orbital excitations as shown in Fig. \ref{fig-config}. Although we can also construct an effective Hamiltonian including these $A_1$ orbitals, we do not try to do this, since the model will become too complicated. Indeed, this simplification is not so bad, since none of the ``ignored" states in Table \ref{tbl-4} is the ground state in any $n_d$ space. It should be noted that the above argument does  not hold near the phase boundary.

\begin{table}[!t]
\caption{Comparison of the model Hamiltonian (\ref{Hamil}) and the results of truncated exact diagonalization for $U=1.5$ eV and $t_{\pi}=-0.085$ eV. $\langle H_{\rm eff}^{\rm 1tet}\rangle/E_{\rm num}$ are shown in the fifth and the sixth columns for $J=0.2$ and $0.3$ eV. The five parameters indicated and a trivial constant term are estimated by using six ``input" states.  The states with $(n_d,S,\Gamma)=(6,1,T_2)$ and $(7,\frac{3}{2},T_1)$ cannot be described by $H_{\rm eff}$ alone.}
\begin{ruledtabular}
  \begin{tabular}{ccccrr}
     $n_d$ & $S$ & $\Gamma$ & $H_{\rm eff}^{\rm 1tet}-C$ & $J=0.2$ eV&$0.3$ eV \\
\hline
     $4$   & $0$ &$A_1$ &   $0$  &  input &input  \\
     $5$   & $\frac{1}{2}$ &$T_2$   & $\epsilon$  & input &input\\
     $6$   & $1$ &$T_1$ &   $2\epsilon+\tilde{U}'-\tilde{J}$  & input &input\\
     $6$   & $0$ &$A_1$ &   $2\epsilon+\tilde{U}+2\tilde{T}$  & input &input\\
     $6$   & $1$ &$T_2$ &   *  & * & *\\
     $6$   & $0$ &$E$ &   $2\epsilon+\tilde{U}-\tilde{T}$  & input &input\\
     $6$   & $0$ &$T_2$ &   $2\epsilon+\tilde{U}'+\tilde{J}$  & input &input\\
     $7$   & $\frac{3}{2}$ &$A_2$   & $3\epsilon+3\tilde{U}'-3\tilde{J}$  & 0.9989 & 1.001\\
     $7$   & $\frac{3}{2}$ &$T_1$   & *  & * & *\\
     $7$   & $\frac{1}{2}$ &$T_2$   & $3\epsilon+2\tilde{U}'+\tilde{U}-\tilde{J}+\tilde{T}$  & 0.9972 & 0.9999 \\
     $7$   & $\frac{1}{2}$ &$E$   & $3\epsilon+3\tilde{U}'$  & 0.9967 & 0.9963 \\
     $7$   & $\frac{1}{2}$ &$T_1$   & $3\epsilon+2\tilde{U}'+\tilde{U}-\tilde{J}-\tilde{T}$  & 0.9974  &0.9980\\
\hline
$C$ &       (eV)&&&$-5.490$ &$-4.861$\\
$\epsilon$ &(eV)&&&$-0.4090$ & $-0.3332$\\
$\tilde{U}$&(eV)&&&$0.3140$ &$0.2626$\\
$\tilde{U}'$&(eV)&&&$0.2989$ & $0.2502$\\
$\tilde{J}$ &(eV)&&&$0.0215$ & $0.0257$\\
$\tilde{T}$ &(eV)&&&$-0.0036$& $-0.017$
        \end{tabular}
\end{ruledtabular}
\label{tbl-4}
\end{table}

\section{\label{sec:4tet}EFFECTIVE MODEL OF FOUR TETRAHEDRA AND CORRELATIONS OF SPIN AND ORBITAL}
In this section, we will construct an effective Hamiltonian describing interacting tetrahedron units in the $^3T_1$ phase. We will then calculate its low-energy eigenstates for the unit of four tetrahedra and the spin and orbital correlation functions for the ground states.

\subsection{Effective Hamiltonian for coupled tetrahedra}\label{HeffCoupledTet}
In Sec. \ref{sec:1tetHeff}, we have constructed an effective model for an isolated tetrahedron unit. We now derive an effective model for coupled tetrahedra in the $^3T_1$ phase by including inter-tetrahedron processes. It is inter-tetrahedron d-electron hoppings that couple otherwise isolated tetrahedron units. 

In Sec. \ref{sec-EigenEnergy}, we obtained low-energy eigenstates $\{|\lambda\rangle\}$ in a tetrahedron unit. When tetrahedron units are decoupled, eigenstates of the whole system are simply direct products of the tetrahedron eigenstates: $|\lambda_1\lambda_2\cdots \lambda_N\rangle$, where $N$ is the number of tetrahedron units. The next step of the real-space renormalization group procedure is to obtain effective couplings between these low-energy states. These tetrahedra are coupled by d-electron hoppings between nearest-neighbor pairs of original sites
\begin{eqnarray}
  H_{\rm hopp}=\sum_{\langle\langle{\bf i},\ {\bf j}\rangle\rangle}\sum_{\sigma\alpha\beta}\Big(t_{{\bf i}{\bf j}}^{\alpha\beta}d_{{\bf i}\alpha\sigma}^{\dagger}d_{{\bf j}\beta\sigma}+{\rm h.c.}\Big),
\end{eqnarray}
where $\langle\langle{\bf i},\ {\bf j}\rangle\rangle$ indicates that $\bf i$ and $\bf j$ are the nearest-neighbor vanadium sites and belong to different unit cells (tetrahedra). There, we need the matrix element of electron hopping processes in the tetrahedron basis $t_{\lambda_{n}\lambda_{m}}^{\lambda'_{n}\lambda'_{m}}$:
\begin{eqnarray}
t_{\lambda_{n}\lambda_{m}}^{\lambda'_{n}\lambda'_{m}}\equiv \langle \lambda_{n}\lambda_{m} |H_{\rm hopp}|\lambda'_{n}\lambda'_{m}\rangle. \label{H11}
\end{eqnarray}
Here, $|\lambda_{n}\lambda_{m} \rangle$ is a direct product state of two tetrahedra $n$ and $m$.
 In practice, we need to calculate the matrix element of the d-electron creation (annihilation) operator $d_{{\bf i}\alpha\sigma}^{\dagger}(d_{{\bf j}\beta\sigma})$. A typical term in Eq. (\ref{H11}) is
\begin{eqnarray}
&&\langle \lambda_{n}\lambda_{m}|t_{{\bf i}{\bf j}}^{\alpha\beta}d_{{\bf i}\alpha\sigma}^{\dagger}d_{{\bf j}\beta\sigma}|\lambda'_{n}\lambda'_{m}\rangle\nonumber\\
&=&t_{{\bf i}{\bf j}}^{\alpha\beta}\langle \lambda_{m}|\langle\lambda_{n}|d_{{\bf i}\alpha\sigma}^{\dagger}d_{{\bf j}\beta\sigma}|\lambda'_{n}\rangle|\lambda'_{m}\rangle\nonumber\\
&=&t_{{\bf i}{\bf j}}^{\alpha\beta}(-1)^{P_{\lambda'_{n}}}\langle \lambda_{n} |d_{{\bf i}\alpha\sigma}^{\dagger}|\lambda'_{n}\rangle \langle \lambda_{m} |d_{{\bf j}\beta\sigma}|\lambda'_{m}\rangle.\label{tmatrix}
\end{eqnarray}
where $\bf i$ $({\bf j})$ belongs to the tetrahedron $n$($m$) and $P_{\lambda^{\prime}_{n}}$ is the electron number in $|\lambda^{\prime}_{n}\rangle$. Since the matrix elements $\langle \lambda_{n} |d_{{\bf i}\alpha\sigma}^{\dagger}|\lambda'_{n}\rangle$ and $\langle \lambda_{m} |d_{{\bf j}\beta\sigma}|\lambda'_{m}\rangle$ in Eq. (\ref{tmatrix}) are local quantities, we can evaluate them for the wavefunctions obtained in Sec. \ref{sec:Eigen1tet}. Using $t_{\lambda_{n}\lambda_{m}}^{\lambda'_{n}\lambda'_{m}}$ obtained in this way, we can write our effective Hamiltonian $H_{\rm eff}$ as

\begin{eqnarray}
H_{\rm eff} &=& \sum_{n\lambda}\epsilon_{\lambda}|\lambda_n\rangle\langle \lambda_n|\nonumber\\
&&+\sum_{\langle n,m\rangle}\sum_{\lambda_n\lambda_m\lambda'_n\lambda'_m}t_{\lambda_{n}\lambda_{m}}^{\lambda'_{n}\lambda'_{m}}|\lambda_n\lambda_m\rangle\langle\lambda'_n\lambda'_m|.\label{H4}
\end{eqnarray}
Here, $\sum_{\langle {n},{m}\rangle}$ is the summation over nearest-neighbor pairs of tetrahedra and $\epsilon_{\lambda}$ is the energy eigenvalue for one tetrahedron which is independent on $n$.

 In the actual calculations, we take not only the $T_2^{(-)}$ orbitals related to the one tetrahedron effective Hamiltonian (\ref{Heff}) but also other orbitals such as $A_1$. This gives corrections to Eq. (\ref{Heff}). Later in Sec. \ref{sec:Conclusion1}, we will further simplify this effective model (\ref{H4}) to a more physical form. Since the matrix element $\langle \lambda_{n} |d_{{\bf i}\alpha\sigma}^{\dagger}|\lambda'_{n}\rangle$ is typically of the order of $\sim 0.3$, and the largest hopping term is $|t_{\sigma}|=0.527$ eV in our calculations, the order of magnitude of $|t_{\lambda_{n}\lambda_{m}}^{\lambda'_{n}\lambda'_{m}}|$ is estimated as $\bar{t}_{\rm eff}\sim(0.3)^2\times0.5=0.045$ eV, the order of $(1/10)|t_{\sigma}|$. This value is relatively smaller than the charge excitation energy of one tetrahedron $\Delta_c\sim 0.1$ eV shown in Fig. \ref{fig-2chiC}. Correspondingly, the exchange interaction among tetrahedron units are of the order of $\bar{t}_{\rm eff}^2/(2\Delta_c)\sim(0.05)^2/(2\times 0.1)=0.0125$ eV. This is a new energy scale of the low-energy properties of this system. The exchange interactions among tetrahedron units will be discussed in Sec. \ref{sec:exchangeTetra}. Indeed, the values of the various exchange interactions turn out to be less than $0.01$ eV.

Before starting the detailed analysis of this model, let us briefly estimate the number of basis states we need to keep for this effective Hamiltonian from the viewpoint of entropy. We are primarily interested in the low-temperature behaviors of LiV$_2$O$_4$ below the coherence temperature $T^*\sim 30$ K. For example the entropy at around $100$ K is ${\mathcal S}(T\simeq 100 {\rm K})\simeq 5k_B\log 2\simeq k_B\log32$ per four vanadium sites determined from the specific heat data.\cite{3} The effective Hamiltonian should have enough degrees of freedom for reproducing this value. 

As discussed in Sec. \ref{sec:1tetHeff}, the average electron density implies that the charge subspaces of $n_d=5,6$ and 7 are dominant local configurations, and it is natural to consider a few lowest-energy states in each subspace. Here we consider the states with partially filled $T_2^{(-)}$ orbitals and count the total entropy per tetrahedron. For these configurations, there are $m_5=6$ states in $n_d=5$ space, $m_6=15$ states in $n_d=6$ space and $m_7=20$ states in $n_d=7$ space. This restricted Hilbert space corresponds to that of Eq. (\ref{Heff}) in the case of one tetrahedron.  Since these three subspaces have different numbers of states, the probability of finding each charge configuration, $P_n$, is not the same to each other but a function of $m$'s:
\begin{eqnarray}
 P_5&=&P_7=(2+\alpha)^{-1}\equiv p,\nonumber\\
 P_6&=&1-2p=(2\alpha^{-1}+1)^{-1},
 \end{eqnarray} 
where $\alpha=\sqrt{m_5m_7/m^2_6}$. The mixing entropy is therefore given by
\begin{eqnarray}
{\mathcal S}_m/k_B&=&-\sum_{n=5}^7 P_n\log P_n\nonumber\\
       &=&-(1-2p)\log(1-2p)-2p\log p.
\end{eqnarray}
Adding the contributions from the degeneracy in each charge subspace, the total entropy is obtained as
\begin{eqnarray}
{\mathcal S}_{\rm tot}&=&{\mathcal S}_m+k_B\sum_{n=5}^7 P_n\log m_n\nonumber\\
           &=&k_B\Big[\log m_6 +\log(1+2\alpha)\Big]\nonumber\\
           &=&k_B\log(m_6+2\sqrt{m_5m_7})\simeq k_B\log36.9.
\end{eqnarray}
 This value is close to the experimental estimate at around $100$ K.\cite{3}

We repeat the same calculation with retaining only the ground states in each
 charge subspace: $m_5=6$, $m_6=9$ and $m_7=4$. This is a minimal set for describing charge fluctuations and electron itineracy. Using the same formula, we obtain this time $\mathcal S\simeq k_B\log (9+4\sqrt{6})\simeq k_B\log 18.8$. This value is  now large enough to reproduce the value at the coherence temperature $T=T^*$: $\sim 2.5k_B \log 2\simeq k_B\log 5.66$. This suggests constructing minimal low-energy effective model defined in this restricted Hilbert space. We can expect that this describes low-energy heavy fermion behaviors. We will propose such a $t$-$J$ like effective model later in Sec. \ref{sec:Conclusion1}.

\subsection{Ground state of four tetrahedron units}\label{4tetPhasediagram}

Now we investigate the ground state when four tetrahedra in $^3T_1$ phase are coupled by electron hoppings. We are primarily interested in the case of $24$ electrons in the four tetrahedra in total. To this end, we employ an exact diagonalization method for the effective model (\ref{H4}). Because of memory limit of our computer, we cut off the high energy states in our diagonalization. We first retain states with $n_d=5,6,$ and $7$ in each tetrahedron, which are minimal states to describe charge fluctuations and thus the inter-tetrahedron superexchange interactions. Secondly, in each $n_d$ subspace, we retain several lowest-energy states only. The number of the retained states $N_{\rm cut}^{n_d}$ is tuned depending on the parameters in the model, and typically $N_{\rm cut}^5=8$, $N_{\rm cut}^6=24$ and $N_{\rm cut}^7=32$.  Inter-tetrahedron correlations are included in the energy level $\epsilon_{\lambda}$ and wavefunctions \{$|\lambda\rangle$\}. The ground-state wavefunction is to be obtained as
\begin{eqnarray}
|g\rangle = {\sum_{\lambda_1}}'{\sum_{\lambda_2}}'{\sum_{\lambda_3}}'{\sum_{\lambda_4}}' W^g_{\lambda_1\lambda_2\lambda_3\lambda_4}|\lambda_1\lambda_2\lambda_3\lambda_4\rangle
\end{eqnarray}
where ${\sum}'$ denotes the sum over the space restricted by $N_{\rm cut}^{n_d}$.

 The approximation of truncating high-energy states is controlled by varying $N_{\rm cut}^{n_d}$ and we have checked our results by increasing the number of retained states.

\begin{figure}[b]
  \begin{center}
    \includegraphics[width=.45\textwidth]{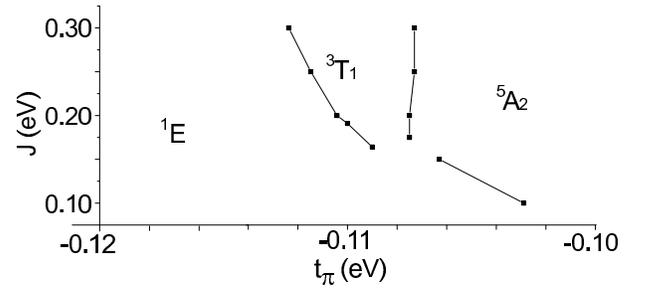}
  \end{center}
\caption{Phase diagram for four tetrahedra with $n_d=24$ and $U=1.5$ eV. $N_{\rm cut}^5=8$, $N_{\rm cut}^6=24$ and $N_{\rm cut}^7=32$. The retained states are those listed in Fig.  \ref{fig-config}.}
\label{fig-4phase}
\end{figure}
First, we show the ground state phase diagram for the case of 24 electrons in four tetrahedra. We calculate ground states by setting $N_{\rm cut}^5=8$, $N_{\rm cut}^6=24$ and $N_{\rm cut}^7=32$ for $J\le 0.3$ eV and the determined phase diagram is shown in Fig. \ref{fig-4phase}. There are three phases: $^1\!E$, $^3T_1$ and $^7\!A_2$. Once again, a state with total spin $S$ belonging to $\Gamma$ representation of the $T_d$ point group is denoted by $^{2S+1}\Gamma$. The ground states change from magnetic to non-magnetic one as $|t_{\pi}|$ increases. This point will be explained in Sec. \ref{sec:exchangeTetra} by estimating the exchange interactions between tetrahedron units. We should note that the phase boundaries do not converge yet with increasing the cut-off numbers \{$N_{\rm cut}^{n_d}$\}. This phase diagram shows approximate, rather than precise, locations of level crossing. However, we can learn a few important characters of the ground state of the four coupled tetrahedra. The first point is that in the shown region of the $t_{\pi}$-$J$ parameter space, these three states, $^1\!E,^3T_1,$ and $^5\!A_2$, are the three lowest multiplets and their energy separations are very small. The second point is that the tendency that the states with large spin appear at the small $|t_{\pi}|$ region is robust among the different truncation numbers used.

\begin{figure}[t]
  \begin{center}
    \includegraphics[width=.45\textwidth]{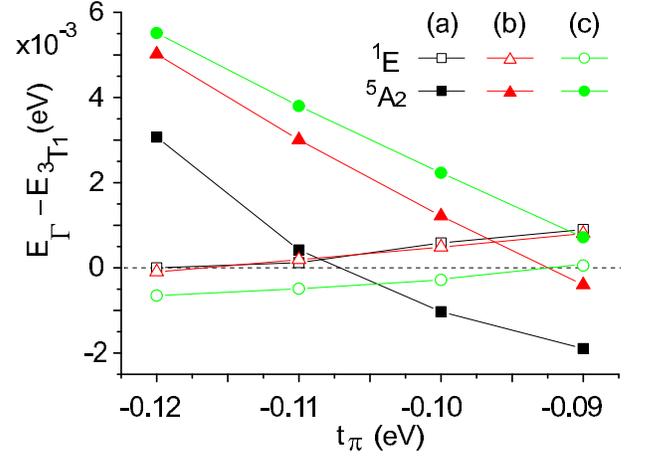}
  \end{center}
\caption{(Color online) $t_{\pi}$ dependence of the ground state energy with $\Gamma$ symmetry $E_{\Gamma}$ relative to that of $^3T_1$ for four tetrahedra calculations for $U=1.5$ eV and $J=0.3$ eV. The results are obtained by three sets of truncation schemes. (a) $(N_{\rm cut}^{5},N_{\rm cut}^{6},N_{\rm cut}^{7})=(8,24,32)$, (b) $(20,34,44)$ and (c) $(32,34,44)$. For $t_{\pi}=-0.12$ eV in (a) $(N_{\rm cut}^{5},N_{\rm cut}^{6},N_{\rm cut}^{7})=(8,34,32)$ is used because of additional near degeneracy.}
\label{fig-result1}
\end{figure}

Figure \ref{fig-result1} shows the $t_{\pi}$ dependence of the energy of the three states appearing in the phase diagram relative to that of $^3T_1$ for different sets of $N_{\rm cut}^{n_d}$, (a) $(N_{\rm cut}^{5},N_{\rm cut}^{6},N_{\rm cut}^{7})=(8,24,32)$, (b) $(20,34,44)$ and (c) $(32,34,44)$. The parameter set (a) is same as that used in Fig. \ref{fig-4phase}. The ground state for (b) is $^1\!E$ for $t_{\pi}=-0.12$ eV, $^3T_1$ for $t_{\pi}=-0.11$ and $-0.10$ eV, and $^5\!A_2$ for $t_{\pi}= -0.09$ eV, and the ground state for (c) is $^1\!E$ for $t_{\pi}\le -0.10$ eV and $^3T_1$ for $t_{\pi}= -0.09$ eV. The energy of $^5\!A_2$ strongly depends on $t_{\pi}$ compared to that of $^1\!E$ and $^3T_1$. In the region of large $|t_{\pi}|$, states with a large spin are energetically unfavored. 
Superexchange via higher-energy virtual states are also present, and some of them generate antiferromagnetic correlations. This is understood by observing the region of the $^5\!A_2$ state shifts to the small $|t_{\pi}|$ region as we increase \{$N_{\rm cut}$\}. It is noted that the $^3T_1$ and $^1\!E$ states are almost degenerate in a wide range of parameters and it is not conclusive which is the ground state within the present calculations. The energy difference between the ground state and the first excited states is typically $10^{-4}$-$10^{-3}$ eV. This might mean the existence of very low-lying excited states in the limit of large $N_{\rm cut}^{n_d}$.

\subsection{Short range correlations}

%

Next, we calculate the spin-spin correlation function $S({\bf q})$ for the sixteen sites in the unit of four tetrahedra. Here, $S({\bf q})$ is an equal-time correlation, i.e., a frequency integrated quantity, and defined by
\begin{eqnarray}
S({\bf q})&=&\frac{1}{N_s}\sum_{{\bf ij}g}\frac{\langle g|S^z_{\bf i}S^z_{\bf j}|g\rangle}{N_g} \exp(i{\bf q}\cdot ({\bf x}_{{\bf i}}-{\bf x}_{{\bf j}})),\\
S^z_{\bf i}&=&\frac{1}{2}\sum_{\sigma\alpha}\sigma n_{{\bf i}\alpha\sigma},
\end{eqnarray}
where $|g \rangle$ and $N_g$ means the index and degeneracy of ground states, respectively. $N_s$ is the number of lattice sites ($N_s=16$ in the present case) and ${\bf x_i}$ is the position of site $\bf i$. Note that $S^z_{\bf i}$ is the spin operator not of a tetrahedron unit but at the vanadium site ${\bf i}$. Since the sum over the ground state degeneracy also includes the spin multiplet, the correlation is of the scalar part of two spin product:
\begin{eqnarray}
\sum_{g}\langle g|S^z_{\bf i}S^z_{\bf j}|g\rangle=\frac{1}{3}\sum_g\langle g|{\bf S}_{\bf i}\cdot{\bf S}_{\bf j}|g\rangle.
\end{eqnarray}
 The matrix element $\langle g|S^z_{\bf i}S^z_{\bf j}|g\rangle$ is calculated by inserting ${\sum}'_{\lambda_1\lambda_2\lambda_3\lambda_4}|\lambda_1\lambda_2\lambda_3\lambda_4\rangle\langle\lambda_1\lambda_2\lambda_3\lambda_4|$ between $S^z_{\bf i}$ and $S^z_{\bf j}$ as a usual procedure. Then, we calculate $\langle g|S^z_{\bf i}S^z_{\bf j}|g\rangle$ from one-tetrahedron matrix elements $\langle \lambda'_n|S^z_{\bf i}|\lambda_n\rangle$ for ${\bf i}\in n $ and the wavefunction of the ground state $|g\rangle$. Spin correlation between the two sites in different tetrahedra and that in the same tetrahedron are given as follows:
\begin{eqnarray}
\langle g|S^z_{\bf i}S^z_{\bf j}|g\rangle&=&{\sum_{\lambda_1\lambda_2}}'{\sum_{\lambda_3\lambda_4}}'{\sum_{\lambda'_3\lambda'_4}}'W^g_{\lambda_1\lambda_2\lambda_3\lambda_4}W^g_{\lambda_1\lambda_2\lambda'_3\lambda'_4}\nonumber\\
&\times&\langle \lambda_3|S^z_{\bf i}|\lambda'_3\rangle\langle \lambda_4|S^z_{\bf j}|\lambda'_4\rangle\ \ \ {\rm for}\ {\bf i}\in 3 \ {\rm and} \  {\bf j}\in 4,\nonumber\\
\label{SzSz1}\\
\langle g|S^z_{\bf i}S^z_{\bf j}|g\rangle&=&{\sum_{\lambda_1\lambda_2\lambda_3}}'{\sum_{\lambda_4\lambda'_4}}'W^g_{\lambda_1\lambda_2\lambda_3\lambda_4}W^g_{\lambda_1\lambda_2\lambda_3\lambda'_4}\nonumber\\
&\times&\langle \lambda_4|S^z_{\bf i}S^z_{\bf j}|\lambda'_4\rangle\ \ \ {\rm for}\ {\bf i}\in 4 \ {\rm and} \  {\bf j}\in 4.\label{SzSz2}
\end{eqnarray}
Here, we have taken the wavefunction $W^g_{\lambda_1\lambda_2\lambda_3\lambda_4}$ as real.

\begin{figure}
  \begin{center}
    \includegraphics[width=.5\textwidth]{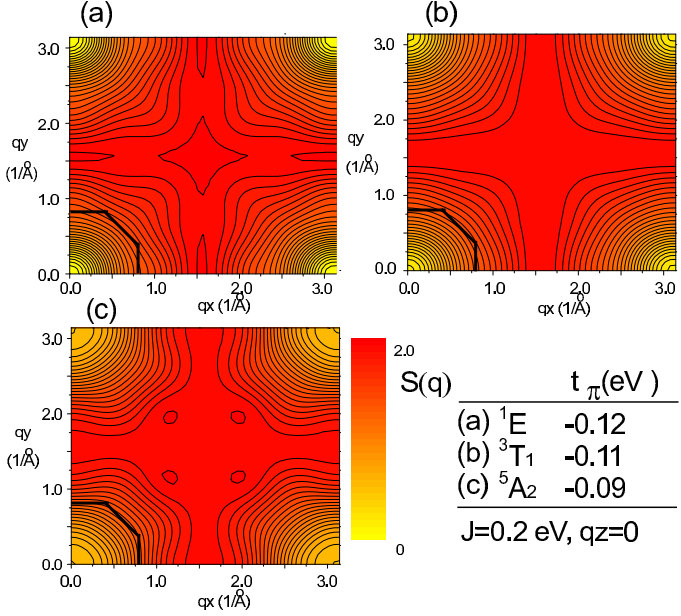} 
  \end{center}
\caption{(Color online) Spin-spin correlation function $S({\bf q})$ on $(q_x,q_y,0)$ plane. $U=1.5$ eV and $J=0.2$ eV. (a) $^1\!E$ state ($t_{\pi}=-0.12$ eV ). (b) $^3T_1$ state ($t_{\pi}=-0.11$ eV ). (c) $^5\!A_2$ ($t_{\pi}=-0.09$ eV ). The first Brillouin zone is indicated by thick lines. The vanadium-vanadium distance is set as $d_{V-V}=2.85$ \AA.}
\label{fig-result2}
\end{figure}

We show $S({\bf q})$ for $U=1.5$ eV and $J=0.2$ eV in Fig. \ref{fig-result2} for the three different phases. This is calculated with the cutoff numbers $(N_{\rm cut}^{5},N_{\rm cut}^{6},N_{\rm cut}^{7})=(8,24,32)$. Since $S({\bf 0})$ is proportional to the ground state expectation value of $(\sum_{\bf i}S^z_{\bf i})^2$, There are notable differences near ${\bf q}={\bf 0}$ for different ground states. There is also difference in the spatial anisotropy in large $\bf q$. In the present calculations, $S(\bf q)$ monotonically increases from the zone center to the zone boundary in all the three ground states. This behavior is different from the finite $|{\bf Q}^*|\sim 0.6$ \AA$^{-1}$ spin fluctuation observed in the neutron experiment.\cite{10} This might be due to the fact that the present $S(\bf q)$ is a frequency integrated quantity, while the neutron experiment observed a low-energy part of spin fluctuations (0.2-0.8 meV).\cite{10}

\begin{figure}[t]
  \begin{center}
    \includegraphics[width=.5\textwidth]{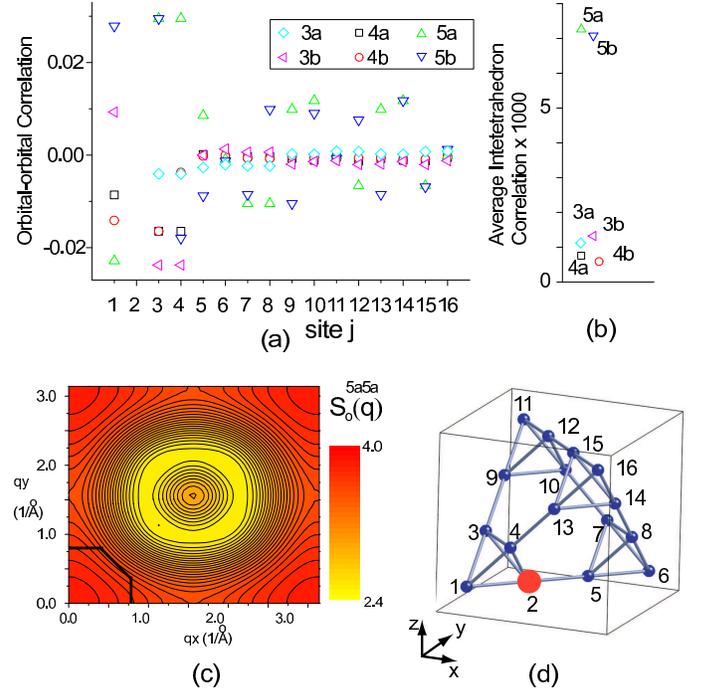}
  \end{center}
\caption{(Color online) Orbital-orbital correlation function in the $\!^1E$ state. $t_{\pi}=-0.12$ eV, $J=0.2$ eV and $U=1.5$ eV. (a) Real space orbital-orbital correlation functions $M^{\alpha\alpha}({2\bf j})$ between site 2 and another site $\bf j$. The site indices are indicated in (d). (b) Average values of inter-tetrahedron correlations defined by $\sum_{{\bf j}=5}^{16}| M^{\alpha\alpha}({2\bf j})|/12$. (c) $S_o^{5a5a}({\bf q})$ on $(q_x,q_y,0)$ plane.}
\label{fig-result3}
\end{figure}

%
We also calculate the orbital correlations. The orbital-orbital correlation function $S_o^{\alpha\beta}({\bf q})$ is defined by
\begin{eqnarray}
S^{\alpha\beta}_{o}({\bf q})&=&\frac{1}{N_s}\sum_{{\bf ij}g}\frac{\langle g|O^{\dagger}_{\alpha}({\bf i})O_{\beta}({\bf j})|g\rangle}{N_g},\nonumber\\
&\times& \exp(i{\bf q}\cdot ({\bf x}_{{\bf i}}-{\bf x}_{{\bf j}})),
\end{eqnarray}
 where $O_{\alpha}$ are orbital operators defined by $O_{4a}=i(d_{yz\sigma}^{\dagger}d_{zx\sigma}-{\rm h.c.})/2$, $O_{4b}=i(d_{zx\sigma}^{\dagger}d_{xy\sigma}-{\rm h.c.})/2$, $O_{4c}=i(d_{xy\sigma}^{\dagger}d_{yz\sigma}-{\rm h.c.})/2$, $O_{5a}=(d_{yz\sigma}^{\dagger}d_{zx\sigma}+{\rm h.c.})/2$, $O_{5b}=(d_{zx\sigma}^{\dagger}d_{xy\sigma}+{\rm h.c.})/2$, $O_{5c}=(d_{xy\sigma}^{\dagger}d_{yz\sigma}+{\rm h.c.})/2$, $O_{3a}=(2d_{xy\sigma}^{\dagger}d_{xy\sigma}-d_{yz\sigma}^{\dagger}d_{yz\sigma}-d_{zx\sigma}^{\dagger}d_{zx\sigma})/\sqrt{12}$, and $O_{3b}=(d_{yz\sigma}^{\dagger}d_{yz\sigma}-d_{zx\sigma}^{\dagger}d_{zx\sigma})/2$ (here the vanadium site {\bf i} and $\sigma$ summation are not shown explicitly). 
The evaluation of the matrix element $\langle g|O_{\alpha}^{\dagger}({\bf i})O_{\beta}({\bf j}) |g\rangle$ is similar to the case of $S(\bf q)$. We show for $^1\!E$ ground states the real space orbital correlations $M^{\alpha\alpha}({\bf ij})\equiv \sum_g\langle g|O_{\alpha}^{\dagger}({\bf i})O_{\beta}({\bf j}) |g\rangle/N_g$ with ${\bf i}=2$ fixed and its Fourier transform $S_o^{\alpha\alpha}({\bf q})$ in Figs. \ref{fig-result3} (a) and (c), respectively. Note that $M^{4b4b}({\bf 2j}) $ and $M^{5b5b}({\bf 2j})$ is identical to $M^{4c4c}({\bf 2j}')$ and $M^{5c5c}({\bf 2j}')$, respectively, where ${\bf j}'$ is the mirror image point of ${\bf j}$ with respect to $(1\bar{1}0)$ plane, and therefore we do not plot the latter. As we can see in Fig. \ref{fig-result3} (a), inter-tetrahedron correlations are strong for the $O_{5a},\ O_{5b}$ and $O_{5c}$ components. This is clearly seen as a difference in the average of $|M^{\alpha\alpha}({2\bf j})|$ for $5\le {\bf j}\le 16$ as shown in Fig. \ref{fig-result3} (b). As for the wavevector dependence, $S^{5a5a}({\bf q})$ has a peak correspondingly at $\bf q=0$ as shown in Fig. \ref{fig-result3} (c). The other modes of orbital fluctuations have similar $\bf{q}$-dependence within the first Brillouin zone. We find similar $M^{\alpha\alpha}(\bf 2j)$ for other ground states and the values of correlations coincide with each other in less than five percents. This means that $^1\!E$, $^3T_1$ and $^5\!A_2$ states have very similar orbital fluctuations but the spin correlations are different as shown in Fig. \ref{fig-result2}.

%

\section{EXCHANGE INTERACTION BETWEEN TETRAHEDRA: SPIN-ORBITAL MODEL}\label{sec:exchangeTetra}
In this section, we will carry out the second order perturbation calculations in the hopping terms and derive a model of Kugel-Khomskii type\cite{KugelKomskii} for the spin $S=1$ and the orbital triplet ($\Gamma=$$T_1$) degrees of freedom, in order to investigate the ground states in more detail. The phase diagram of four tetrahedra obtained in Sec. \ref{4tetPhasediagram} will be explained in terms of various exchange interactions such as pure magnetic, pure orbital, and coupled magnetic and orbital exchange interactions. Characteristic orbital configurations coupled to spin degrees of freedom in terms of tetrahedron units will be discussed in the final part of this section.

\subsection{Exchange Hamiltonian}

In order to uncover the obtained ground states and their properties, we investigate various exchange interactions of spin and orbital degrees of freedom between different tetrahedra. To examine orbital and spin correlations, we temporarily neglect charge fluctuations and consider $^3T_1$ multiplet in $n_d=6$ space at each tetrahedron. As in a usual manner, we carry out a calculation of the second order perturbation in the hopping terms (the last term in Hamiltonian (\ref{H4})), and derive Kugel-Khomskii type exchange interactions\cite{KugelKomskii} of spin and orbital degrees of freedom. In the second order perturbations, nine states of $^3T_1$ multiplet are used as initial and final states, while eight states in $n_d=5$ and thirty-two states in $n_d=7$ are kept as virtual states. For $n_d=5$ and $7$ states, we keep states with $S=1/2$ and $3/2$ in the low-energy spectra, since the unperturbed states are those with $S=1$ for $n_d=6$.

\begin{figure}[t!]
  \begin{center}
    \includegraphics[width=.5\textwidth]{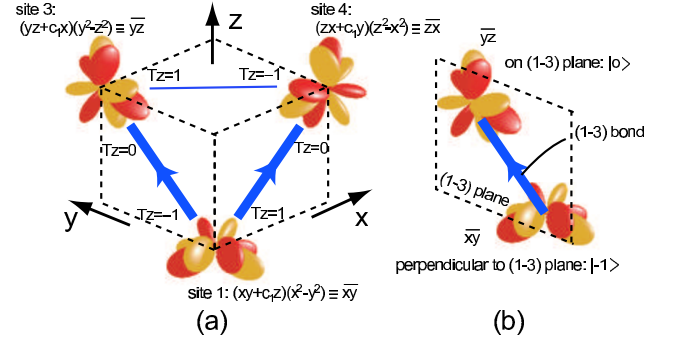}
  \end{center}
\caption{(Color online) (a) An example of orbital configuration. Wavefunctions of $T_1$ are graphically drawn with $c_1=0.4$. The site indices correspond to those in Fig. \ref{fig-lat} (b) Thick line with arrow indicates the bond favored by $J_1^{88}$. The direction of the arrow corresponds to that in the graph in Fig. \ref{fig-closetriangle}. (b) Details of the definition $T_z$ for bond (1-3).}
\label{fig-help}
\end{figure}

 We assign the state whose orbital is on the plane\cite{explaneplane} including the bond ($n$-$m$) as $T_z=0$ one ($\equiv |0\rangle$). The other two states are assigned to $T_z=\pm $ ($\equiv |\pm\rangle$). We show in Fig. \ref{fig-help} (a) an example of this assignment. We use simplified notations for the orbital label in $T_1$ representations hereafter such as $\overline{xy}\equiv (xy+c_1z)(x^2-y^2)$ and so on. It should be noted that the same orbital is assigned to different $T_z$-states depending on bond directions, as depicted. We use for the orbital part eight operators, $T_{\mu}$ $(\mu=1,2,\cdots,$ and $8)$. For orbital degrees of freedom, we introduce a representation that depends on the bond direction. Let us consider a bond and orbitals at the ends of it. $\mu=1,2$ and $3$ correspond to the pseudospin-1 operator $T_x,\ T_y$, and $T_z$, respectively. For $\mu \ge 4$, we define $T_4\equiv \{T_x, T_y\}$, $T_5\equiv \{T_y, T_z\}$, $T_6\equiv \{T_z, T_x\}$, $T_7\equiv T_x^2-T_y^2$, and $T_8\equiv (2T_z^2-T_x^2-T_y^2)/\sqrt{3}$. For the spin part, we use standard spin-1 operators $S_a$ $(a=x,y,$ and $z)$. Using these operators, the exchange Hamiltonian between tetrahedra $n$ and $m$ (bond ($n$-$m$)) reads
\begin{eqnarray}
H_{\rm ex}^{nm}&=&\sum_{\mu,\nu=0}^8\Big\{ \Big[\frac{2}{3}J_1^{\mu\nu}({nm})+J^{\mu\nu}_2({nm}){\bf S}(n)\cdot {\bf S}(m)\Big]\nonumber\\
&\times&T_{\mu}({ n})T_{\nu}({ m})\Big\},\label{ExHamil}
\end{eqnarray}
where $T_{\mu}(n)$ (${\bf S}(n)$) means the orbital (spin) operator at tetrahedron $n$ and $T_0(n)\equiv \sqrt{2/3}$. The pre-factor of $J_1^{\mu\nu}$ is just the normalization. $J_1^{00}$ is nothing but the origin of energy and we set $J_1^{00}=0$.

Due to the symmetry of the $T_1$ orbital, selection rules exist for $J_1^{\mu\nu}$ and $J_2^{\mu\nu}$ and some elements vanish. There are two types of symmetry operations which are used to reduce the number of independent coupling constants. (i) mirror: $|+\rangle\leftrightarrow|-\rangle$ for both $n$ and $m$ sites simultaneously, and (ii) $C_2$ rotation: $n\leftrightarrow m$. First, under the operation (i), operators $T_3$, $T_5$ and $T_6$ at each tetrahedron change their sign while the others do not. Thus, the products including one of the former group, for example, $T_3(n)T_7(m)$, cannot appear in the exchange interactions, therefore $J_1^{37}(nm)=0$. Secondly, under the operation (ii), $T_{\mu}(n)$ ($\mu=$1,2,5, and 6) are transformed to $-T_{\mu}(m)$. The others change their site index but do not change their sign. From this, the terms including one of them appear in antisymmetric combination, $T_{\mu}(n)T_{\nu}(m)-T_{\nu}(n)T_{\mu}(m)$. The other terms appear in symmetric combination $T_{\mu}(n)T_{\nu}(m)+T_{\nu}(n)T_{\mu}(m)$. This leads to $J_1^{\mu\nu}=-J_1^{\nu\mu}$ for the first case, while $J_1^{\mu\nu}=J_1^{\nu\mu}$ for the second case. Using these properties and the fact that the exchange interactions are real\cite{imag}, it turns out that the number of independent couplings is 13 in $J_1^{\mu\nu}$ and 21 in $J_2^{\mu\nu}$\cite{reduction}.

%

\begin{figure}[t!]
  \begin{center}
    \includegraphics[width=.4\textwidth]{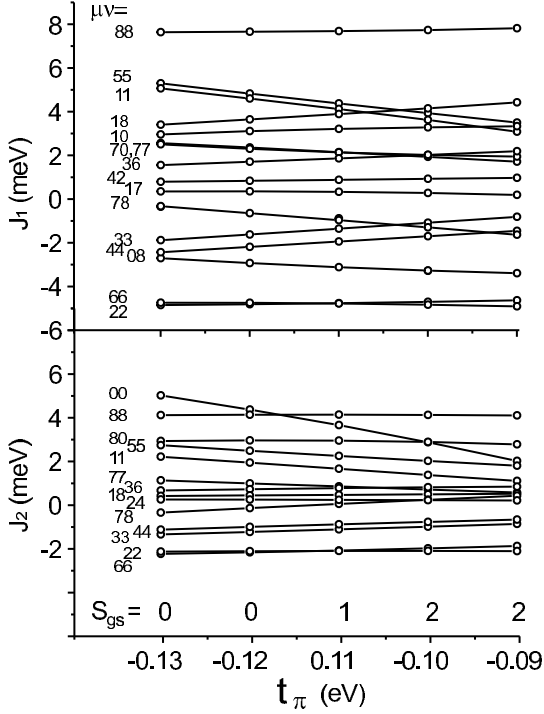}
  \end{center}
\caption{Spin and orbital exchange couplings $J_1^{\mu\nu}(12)$ and $J_2^{\mu\nu}(12)$ as a function of $t_{\pi}$. $U=1.5$ eV, $J=0.2$ eV. The three smallest $J_2^{\mu\nu}(12)$ and $J^{00}_1(12)$ are not shown for simplicity. $S_{\rm gs}$ denotes the ground state spin for each $t_{\pi}$.}
\label{fig-Coupl}
\end{figure}
%

We now calculate $J_1^{\mu\nu}$ and $J_2^{\mu\nu}$ from the numerically obtained $H_{\rm ex}^{nm}$, and to this end we use the following Fierz identities,
\begin{eqnarray}
{\rm Tr}[S_{a}(n)S_{b}(n)]&=&3\times 2\delta_{ab},\label{SScom}\\
{\rm Tr}[T_{\mu}(n)T_{\nu}(n)]&=&3\times2\delta_{\mu\nu}\label{TTcom},
\end{eqnarray}
where Tr is taken over in both spin and orbital spaces of one tetrahedron.
Using Eqs. (\ref{SScom}) and (\ref{TTcom}), we obtain
\begin{eqnarray}
J_1^{\mu\nu}(nm)\!\!&=&\!\!\frac{1}{3\times 2^3}{\rm Tr'}[T_{\mu}({n})T_{\nu}({m})H_{\rm ex}^{{nm}}],\\
J_2^{\mu\nu}(nm)\!\!&=&\!\!\frac{1}{3\times2^4}{\rm Tr'}[{\bf S}(n)\!\cdot\!{\bf S}(m)T_{\mu}({n})T_{\nu}({m})H_{\rm ex}^{{nm}}],
\end{eqnarray}
where ${\rm Tr}'$ is taken over in both spin and orbital spaces for two tetrahedra $n$ and $m$. In Fig. \ref{fig-Coupl}, we show $J_1^{\mu\nu}$ and $J_2^{\mu\nu}$ for the $(1$-$2)$ bond as a function of $t_{\pi}$. Since $J_a^{\mu\nu}=\pm J_a^{\nu\mu}$ for $a=1$ and $2$, we plot only one of them. It is found that all couplings are smaller than $10$ meV, which is consistent with experimental results for the Weiss temperature $\Theta'\sim -40$-$-30$ K estimated below 400 K\cite{1,8,9}.

The coupling $J_2^{00}$ is pure spin exchange and decreases with decreasing $|t_{\pi}|$, indicating the enhancement of ferromagnetic processes. We can explain this tendency by examining important virtual processes. When adding an electron of $A_1$ orbital to the $^3T_1^{n_d=6}$ ground state, we obtain basically $^4T_1^{n_d=7}$ state which is the lowest excited state in $n_d=7$ subspace (see Fig. \ref{fig-config} and Table \ref{tbl-4}). Since this virtual state $^4T_1^{n_d=7}$ has spin $3/2$, $^4T_1^{n_d=7}$ state contributes to ferromagnetic exchange interactions in the second order perturbations. The point is that the energy of this state decreases as $|t_{\pi}|$ decreases. Therefore, the ferromagnetic interactions are enhanced. As $|t_{\pi}|$ increases, the energy of $^4T_1^{n_d=7}$ state increases. In the large $|t_{\pi}|$ region, the antiferromagnetic exchange interactions generated via the excited $^2T_2^{n_d=7}$, $^2E^{n_d=7}$ and $^2T_1^{n_d=7}$ states dominate. As a result, $J_2^{00}(nm){\bf S}(n)\cdot{\bf S}{(m)}$ (pure magnetic exchange interaction) notably becomes strong among others and this is antiferromagnetic coupling. Thus, the inter-tetrahedron exchange interaction depends significantly on the excitation energy of $A_1$ orbital. This also explains the tendency observed in Fig. \ref{fig-4phase}, i.e., magnetic phases appear in small $|t_{\pi}|$ regions.

\subsection{Spin-orbital model: four coupled tetrahedron units}
\begin{figure}[b!]
  \begin{center}
    \includegraphics[width=.4\textwidth]{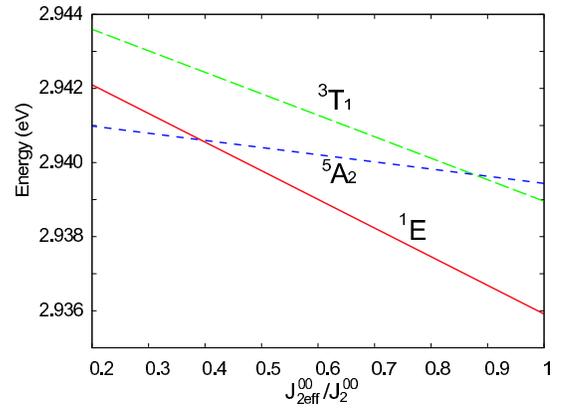}
  \end{center}
\caption{(Color online) The lowest three energy eigenvalues vs $J_{2{\rm eff}}^{00}/J_2^{00}$. $t_{\pi}=-0.10$ eV and the other parameters are the same as those in Fig.  \ref{fig-Coupl}.}
\label{fig-J2eff00}
\end{figure}

With the obtained couplings $J_1^{\mu\nu}$ and $J_2^{\mu\nu}$, we numerically diagonalize the spin-orbital exchange model for the coupled four-tetrahedron system,
\begin{eqnarray}
H_{\rm ex}=\sum_{1\le n<m\le 4}H_{\rm ex}^{nm},\label{FullEX}
\end{eqnarray}
and calculate a few lowest-energy states. The result is that the ground state is $^1\!E$ state for $J=0.2$ eV and $-0.13\le t_{\pi}\le -0.09$ eV. By comparing this result to the phase diagram of Fig. \ref{fig-4phase}, it turns out that the perturbative calculations underestimate the ferromagnetic exchange coupling as is easily understood by observing the lack of double-exchange interactions in Hamiltonian (\ref{FullEX}). The results of the present perturbative analysis is similar to the cut-off scheme (c) in Fig. \ref{fig-result1} except $t_{\pi}=-0.09$ eV where the ground state is $^3T_1$.  Then, in order to check whether we can explain the phase diagram in Fig. \ref{fig-4phase} by the exchange model (\ref{FullEX}), we carry out the same calculation by replacing perturbatively calculated $J_2^{00}$ by $J_{2{\rm eff}}^{00}$. By introducing an effective pure magnetic exchange interaction $J_{2{\rm eff}}^{00}$, we can incorporate enhancement of ferromagnetic correlations. We show the three lowest-energy eigenvalues obtained in this way in Fig. \ref{fig-J2eff00}. The three lowest states are indeed those appearing in the phase diagram Fig. \ref{fig-4phase}. This indicates that the present perturbative calculations capture the essential part of this system. As $J_{2{\rm eff}}^{00}$ decreases, the ground state changes from $^1\!E$ to $^5\!A_2$. However, the $^3T_1$ state does not become the ground state with varying $J_2^{00}$ only, and therefore we would need more complete manipulations of the exchange coupling constants.
%

\subsection{Orbital wavefunction}\label{sec:orbwave}
\begin{figure}[b!]
  \begin{center}
    \includegraphics[width=.4\textwidth]{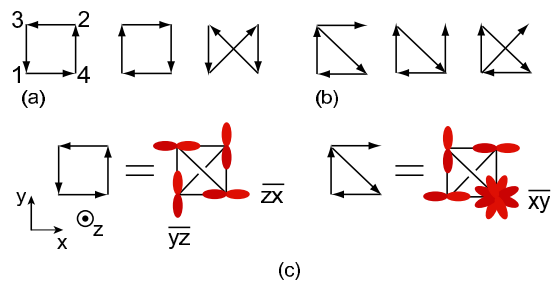}
  \end{center}
\caption{(Color online) Typical graphical representations of orbital states favored by $J_1^{88}$. The site indices are indicated by the numbers $1\sim 4$ around the first graph in (a). Examples of (a) ``closed path'' graphs, and (b) ``lasso'' graphs. (c) Actual orbital shapes in two representatives. }
\label{fig-closetriangle}
\end{figure}

Now we investigate in detail the orbital part of low-energy eigenstates of the exchange model $H_{\rm ex}$ for the coupled four-tetrahedron system. The important point is that among the pure orbital interactions $\{J_1^{\mu\nu}\}$, $J_1^{88}(nm)T_8(n)T_8(m)$ term is always the largest and nearly independent of $t_{\pi}$. Since the orbital operator $T_8$ is defined as
\begin{eqnarray}
T_8=\frac{1}{\sqrt{3}}\big[|+\rangle\langle +| -2|0\rangle\langle 0|+ |-\rangle\langle -|\big],
\end{eqnarray}
this term favors the bond configurations in which one orbital lies on the plane including the bond ($|0\rangle$) and the other does on the plane perpendicular to that ($|\pm \rangle$). The system of four coupled tetrahedra has 30 such states, and four out of the six bonds have the favored configurations in each of them. We can illustrate these 30 states by simple graph representations. Typical graphs are shown in Fig. {\ref{fig-closetriangle}}. Vertices of the square represent tetrahedra. For each bond satisfying the condition above, we draw an arrow which ends at the vertex (tetrahedron) where the orbital state is local $|0\rangle$. In this representation, there is at most one arrow going in a vertex but more than one arrows can go out from a vertex. There are two distinct types of graphs. The graphs in Fig. \ref{fig-closetriangle} (a) are ``closed path'' graphs and contain two orbitals of $T_1$ multiplet. The graphs in Fig. \ref{fig-closetriangle} (b) have a shape similar to lasso (rope with a noose at end) and contain three orbitals of $T_1$ multiplet. 

\begin{figure}[t!]
  \begin{center}
    \includegraphics[width=.4\textwidth]{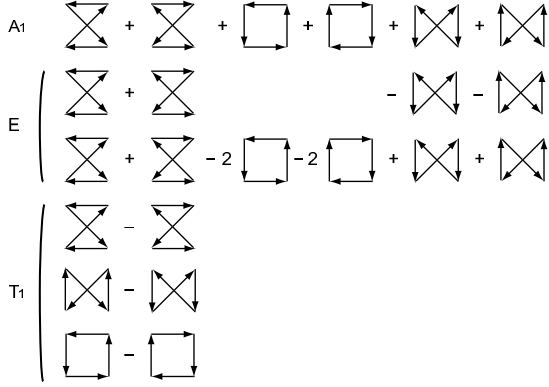}
  \end{center}
\caption{Six orbital basis states of type (a) in Fig. \ref{fig-closetriangle} classified by $T_d$ point group symmetry. Normalization factor is not shown.}
\label{fig-OrbIR}
\end{figure}

The orbital part of the ground states for four tetrahedra can be well described by linear combinations of these 30 orbital states. When setting $J_2^{\mu\nu}=0$, we can show that three lowest-energy orbital eigenstates have $A_1$, $E$, and $T_1$ symmetries. Once again, symmetry classification is useful to understand this. The states of type (a)  are classified as $A_1\oplus E \oplus T_1$, and those of the (b) type are classified as $A_1\oplus A_2\oplus 2E \oplus 3T_1 \oplus 3T_2$. In Fig. \ref{fig-OrbIR}, we graphically show the basis states of each irreducible representation for the type (a). Those for the type (b) are shown in Appendix \ref{AppOrbWavefunc}. The pure orbital terms $J_1^{\mu\nu}$  hybridize type (a) and (b) states. The states of each representation interact in the Hamiltonian only with those of the same representation. Therefore, as far as these 30 orbital states are concerned, the size of the matrix to diagonalize is reduced to, $2,\ 3$, and $4$ for $A_1$, $E$, and $T_1$ representation, respectively, and we can diagonalized them analytically. The matrix elements for this restricted Hilbert space are calculated in Appendix \ref{AppOrbWavefunc}. It is noted that another diagonal interaction $J_1^{33}<0$ lifts the degeneracy of type (a) and (b), and favors type (a) configurations. Taking into account the hybridizations between type (a) and (b) states, the representations appearing in type (a) would have a lower energy. This explains why $A_1$, $E$, and $T_1$ orbital states are the three lowest-energy states.

%

%

\subsection{Spin-orbital coupled wavefunctions}\label{(0E)}
Now let us go back to the phase diagram Fig. \ref{fig-4phase} and discuss these three types of ground states. In the previous subsection, we have discussed the low-energy orbital part in detail. Now we proceed to study the spin part together with the orbital one. As shown above, the low-energy orbital states are linear combinations of the type (a) and (b) states shown in Fig. {\ref{fig-closetriangle}}. For simplicity, we here discuss only the  type (a) configurations, since the weight of type (a) is about two times larger than type (b) in the present parameter sets. The six states of type (a) are reduced to three irreducible representations $A_1^{\rm orb}\oplus E^{\rm orb}\oplus T_1^{\rm orb}$ of $T_d$ point group as shown in Fig. \ref{fig-OrbIR}. Each of the three has only one set of basis states and therefore these states are automatically eigenstates of any orbital Hamiltonian with $T_d$ symmetry as far as the type (a) states are dominant. When the spin-orbital couplings $J_2^{\mu\nu}$ are switched on, these irreducible representations of orbital are to be hybridized to constitute eigenstates of the spin-orbital system $H_{\rm ex}$. 

First, we start to discuss $^1\!E$ states. The $T_d$ point group symmetry of the system implies that the $S=0$ sector of spin wavefunctions in four tetrahedra is decomposed to two irreducible representations $A_1^{\rm spin}\oplus E^{\rm spin}$ as shown in Appendix \ref{AppS=0}. Since the ground state considered now belongs to $E$ representation, this state is a linear combination of $E^{\rm orb}\otimes A_1^{\rm spin}$, $A_1^{\rm orb}\otimes E^{\rm spin}$ and the $E$ representation in $E^{\rm orb}\otimes E^{\rm spin}=A_1\oplus A_2 \oplus E$. Our calculation shows that, among them, $E^{\rm orb}\otimes A^{\rm spin}$ and $E^{\rm orb}\otimes E^{\rm spin}$ components are much larger than that of $A_1^{\rm orb}\otimes E^{\rm spin}$. These dominant two components are entangled with each other, i.e., the wavefunction is not approximated by a single product of spin and orbital parts. This means that the spin and orbital are strongly coupled with each other.

\begin{figure}[t!]
  \begin{center}
    \includegraphics[width=.45\textwidth]{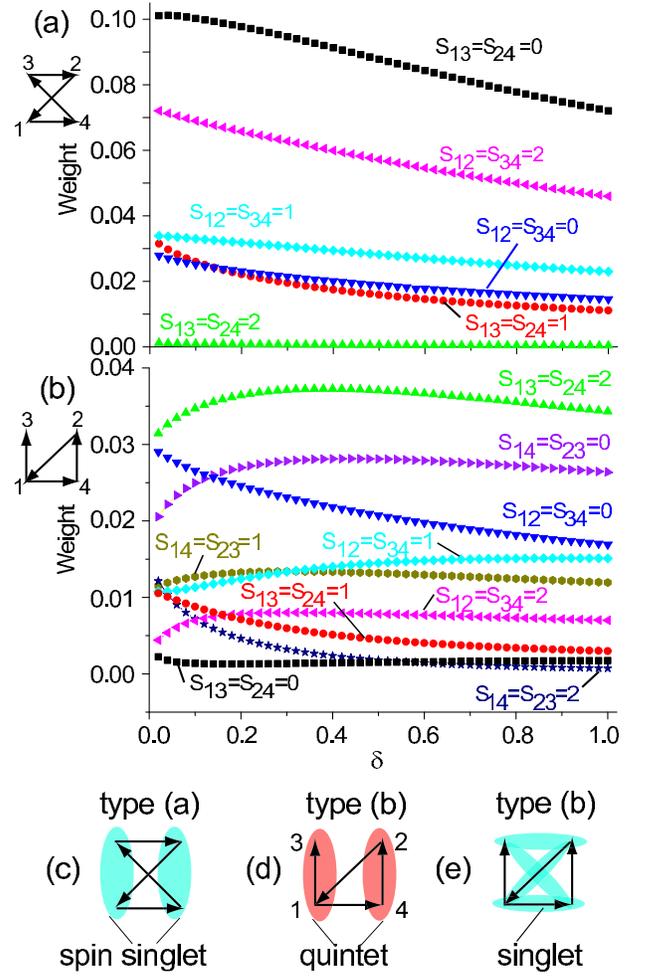}
  \end{center}
\caption{(Color online) Weight of spin wavefunction on each bond vs $\delta$. $U=1.5$ eV, $J=0.2$ eV and $t_{\pi}=-0.12$ eV. (a) Case of ``closed path'' orbital configuration. From the symmetry, Note that $(S_{14},S_{24})$ is equivalent to $(S_{12},S_{34})$. (b) Case of ``lasso'' orbital configuration. (c)-(e) Various spin pair correlations.}
\label{fig-deltaSS}
\end{figure}
%

To discuss the correlations of orbital and spin further, let us calculate for the ground state the probability that, upon fixing the orbital configuration to a given one, the two spins ${\bf S}(n)$ and ${\bf S}(m)$ have the total spin $S_{nm}$. Note that the other two spins also have the same total spin $S_{nm}$, since the $^1\!E$ state is spin singlet. The results for two representative orbital configurations are plotted in Fig. \ref{fig-deltaSS} upon gradually switching on the spin-orbital couplings. Namely, a control parameter $\delta$ is introduced to replace $J_2^{\mu\nu}\to \delta J_2^{\mu\nu}$ with $0\le\delta\le 1$, and the correlations are plotted as a function of $\delta$. If there are no correlations either in orbital or spin parts, the probability is $1/90\sim 0.011$, and the configurations with much larger probability are dominant ones. In the orbital part, each type (a) configuration has weight $0.13$, while  $0.04$ for type (b) at $\delta=0$. Overall difference in weights between Fig. \ref{fig-deltaSS} (a) and (b) is due to this difference in the orbital weights. 

It is noted that spin fluctuations are strongly correlated with orbital configurations. The position of spin-singlet (spin-quintet) tetrahedron pair is correlated with local orbital configurations as shown in Fig \ref{fig-deltaSS} (a) (Fig. \ref{fig-deltaSS} (b)).
\begin{table*}[!t]
\caption{Matrix elements of four dominant exchange interactions for each configuration of two-tetrahedron units: the total spin $S_{12}$ and orbital $(T_z(1),T_z(2))$. $(\pm,\pm)$ is the representative for $(+,+)$, $(+,-)$, $(-,+)$ and $(-,-)$, and $(\pm,0)$ is the representative for $(+,0)$, $(-,0)$, $(0,+)$ and $(0,-)$. }
\begin{ruledtabular}
  \begin{tabular}{ccccccccccc} 
  types of exchange interactions     & $S_{12}$&$2$ & $2$ &$2$ & $1$ &$1$ &$1$ & $0$ & $0$& $0$\\
       & $(T_z(1),T_z(2))$&$(\pm,\pm)$ & $(\pm,0)$ & $(0,0)$ & $(\pm,\pm)$ & $(\pm,0)$ & $(0,0)$& $(\pm,\pm)$ & $(\pm,0)$ & $(0,0)$\\
\hline
$T_8(1)T_8(2)$& &$\frac{1}{3}$& $-\frac{2}{3}$&$\frac{4}{3}$ &$\frac{1}{3}$ &$-\frac{2}{3}$ &$\frac{4}{3}$ &$\frac{1}{3}$ &$-\frac{2}{3}$ &$\frac{4}{3}$\\
${\bf S}(1)\cdot {\bf S}(2)$& &$1$& $1$&$1$ &$-1$ &$-1$ &$-1$ &$-2$ &$-2$ &$-2$\\
${\bf S}(1)\cdot {\bf S}(2)T_8(1)T_8(2)$& &$\frac{1}{3}$& $-\frac{2}{3}$&$\frac{4}{3}$ &$-\frac{1}{3}$ &$\frac{2}{3}$ &$-\frac{4}{3}$ &$-\frac{2}{3}$ &$\frac{4}{3}$ &$-\frac{8}{3}$\\
${\bf S}(1)\cdot {\bf S}(2)[T_8(1)T_0(2)+T_0(1)T_8(2)]$& & $\frac{2\sqrt{2}}{3}$& $-\frac{\sqrt{2}}{3}$&$\frac{4\sqrt{2}}{3}$ &$-\frac{2\sqrt{2}}{3}$ &$\frac{\sqrt{2}}{3}$ &$\frac{4\sqrt{2}}{3}$ &$-\frac{4\sqrt{2}}{3}$ &$\frac{2\sqrt{2}}{3}$ &$\frac{8\sqrt{2}}{3}$
  \end{tabular}
\end{ruledtabular}
\label{tbl-JSSTT}
\end{table*}
For type (a) graphs, spin-singlet correlations are strong in the tetrahedron pair for which the orbital energy is not favored, i.e., bonds without arrow in the figure. This tendency is understood by noting that, next to the largest coupling $J_1^{88}$, the dominant coupling constants are $J_2^{00}$, $J_2^{88}$ and $J_2^{08}$ as seen in Fig. \ref{fig-Coupl}. Their contributions to energy are compared for different spin-orbital configurations in Table \ref{tbl-JSSTT}.  The largest coupling is the pure magnetic exchange $J_2^{00}$ which is antiferromagnetic. The others are spin-orbital couplings $J_2^{88}$ and $J_2^{08}$ which are both positive. Type (a) states have only $(0,\pm)$, $(\pm,0)$ and $\pm(1,1)$ configurations. The sum of the three terms give the lowest energy for spin singlet on the bonds $\pm(1,1)$, i.e., there exist singlet correlations between bonds without arrow. As for the $(0,\pm)$ orbital sector, the energy balance is more delicate but the maximum spin configuration is stabilized. This is because although the pure spin coupling favors the singlet one, the energy gain from $J_2^{80}$ and $J_2^{88}$ is larger for $S_{12}=2$ configuration.

For type (b) graphs, spin-quintet correlations are strong on two of six bonds, (1-3) and (2-4), as shown in Fig. \ref{fig-deltaSS} (d). Since the total spin is singlet, this also means that spin-singlet correlations are enhanced on the other four bonds as depicted in Fig. \ref{fig-deltaSS} (e). Existence of ferromagnetic correlations can be explained as follows. As we discussed in the case of type (a) graphs, the ferromagnetic correlations are enhanced on the bonds $(0,\pm)$ and $(\pm,0)$. Since the total spin is singlet, two quintets should not be overlapped. Combining these implies ferromagnetic spin correlations on the (1-3) and  antiferromagnetic correlations on all the others.

\begin{figure}[t!]
  \begin{center}
    \includegraphics[width=.45\textwidth]{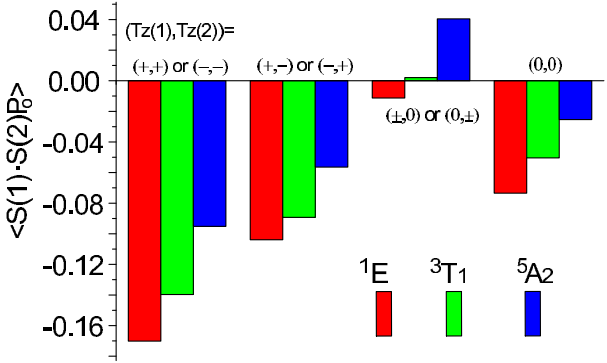}
  \end{center}
\caption{(Color online) Spin-spin correlation function projected to four distinct cases of orbital configurations $\langle {\bf S}(1) \cdot {\bf S}(2)P_o\rangle$ for $^1\!E$, $^2T_1$ and $^5\!A_2$ states. $P_o$ is the orbital projection operator to $(T_z(1),T_z(2))=(\tau,\tau')$ with $\tau,\tau'=\pm$, and $0$. The parameters used are $t_{\pi}=-0.11$ eV, $U=1.5$ eV and $J=0.3$ eV.}
\label{fig-SS}
\end{figure}

The other two ground states, namely, $^3T_1$ and $^5\!A_2$ states can be understood in the same way. The $^5\!A_2$ state has no component of the spin wavefunction with $T_2$ symmetry as shown in Appendix \ref{AppS=2}, proved by the symmetry argument. We show in Fig. \ref{fig-SS} the spin-spin correlation for the three types of ground states. Here, instead of the usual spin-spin correlation, we decompose it into nine parts each of which corresponds to a different orbital configuration on the bond considered. Therefore, the plotted value includes the probability of each orbital configuration. Summing up over the nine parts leads to the ordinary spin-spin correlation. We can see that in both of $^3T_1$ and $^5\!A_2$ states, the spin singlet correlations are strong at $(\pm,\pm)$ orbital configurations as in the $^1\!E$ case. In the state with larger total spin, of course, spin-spin correlation generally becomes more ferromagnetic (shift towards positive). For $^1\!E$ and $^3T_1$, the spin-spin correlations are nearly absent at $(0,\pm)$ orbital configurations. This comes from the bond average in the definition of the spin-spin correlation in Fig. \ref{fig-deltaSS} and ferromagnetic contributions (quintets depicted in Fig. \ref{fig-deltaSS}) almost cancel with antiferromagnetic ones (singlet).

%

\section{\label{sec:Conclusion1}DISCUSSIONS}

Let us reconsider the effective renormalized 
Hamiltonian. It is represented in terms of tetrahedron variables
 as Eq. (\ref{H4}). We first analyze it numerically and further 
discussed its kinetic and interaction terms separately, and 
found several important processes.  Now let us 
assemble these pieces to build a full effective 
Hamiltonian. Since the low-energy physics is concerned, 
electron hoppings are constrained, and among them we keep 
minimal hopping processes. Namely, the electron 
number in each tetrahedron is limited to $n_d$=5, 6, 
or 7, and we consider only the hopping processes 
between ground states in these subspaces.  
The processes including excited states are 
neglected. It is possible to represent this constraint
 if we consider only $T_2^{(-)}$-molecular 
orbital, as discussed in Sec. \ref{sec:shell}. 
The $E$ orbital is fully occupied 
in the cases considered, and we represent this configuration 
 as vacuum. We have shown in 
Sec. \ref{sec:Eigen1tet} that the ground states 
for $n_d$=5, 6, and 7 have spin $S=1/2,\ 1$ and $3/2$, respectively.
This means that constrained electron hoppings generate 
ferromagnetic double exchange interactions. We note that, as we discussed in Sec. \ref{HeffCoupledTet}, the total entropy of this restricted Hilbert space is ${\mathcal S}\sim k_B \log 18.8$ per tetrahedron, which is larger than the experimental value at the coherence temperature $T^*$, $k_B\log 5.66$.\cite{3} Thus, we expect that this restricted Hilbert space has large enough  degrees of freedom to describe the heavy fermion behaviors at low temperatures.

The other terms of the effective 
model are interactions.  We can use the same 
spin-orbital exchange Hamiltonian (\ref{ExHamil}) for this part, 
but need to modify its coupling constants.  
The values of couplings $J_1^{\mu\nu}$ and $J_2^{\mu\nu}$ 
shown in Fig.\ref{fig-Coupl} 
were determined by including all the possible 
hopping processes 
in the second-order perturbation.  
However, now that we treat the hopping processes 
connecting ground states as 
real processes in the kinetic term, 
we need to subtract their 
contributions from the exchange coupling constants.

These arguments show that a key issue is 
the competition between ferro- and antiferro-magnetic
 interactions coupled with orbital degrees of freedom. 
This was discussed in Sec. \ref{(0E)}, and we come to a conclusion
 that the low-energy effective model is similar to the $t$-$J$ model of 
high-temperature superconductor; it is defined in 
terms of a localized spin one with orbital triplet, and mobile
 quasiparticle with $T_2^{(-)}$ symmetry on the effective lattice (f.c.c.).
 The localized spin and orbital
 degrees of freedom are coupled via exchange interactions
 between nearest neighbors. The hopping processes
 of the mobile quasiparticles change local spin and orbital
 configurations. 
The leading terms of the effective Hamiltonian read
\begin{eqnarray}
H_{\rm eff}&=&\sum_n\Big(-\mu_{\rm eff} N(n) +U_{\rm eff}N^2(n)\Big)\nonumber\\
&+&\!\sum_{\langle n,m\rangle}\Big[\sum_{\alpha\beta\sigma}\Big( \tilde{t}_{nm}^{\alpha\beta}Pa_{n\alpha\sigma}^{\dagger}a_{m\beta\sigma}P+{\rm h.c.}\Big)\nonumber\\
&+&\sum_{\mu\nu}\Big(\frac{2}{3}\tilde{J}^{\mu\nu}_1(nm)+\tilde{J}^{\mu\nu}_2(nm){\bf S}(n)\cdot {\bf S}(m)\Big)\nonumber\\
&\times& T_{\mu}(n)T_{\nu}(m)\Big]+\cdots,\label{efftJ}
\end{eqnarray}
 where $P$ is the projection operator to the restricted Hilbert space, namely, the ground states of $n_d=5$, $6,$ and $7$ spaces. $a_{n\alpha\sigma}$ represents the mobile quasiparticle with the $T_{2}^{(-)}$ orbital and the spin $\sigma$ at the tetrahedron $n$. $N(n)$ is the number operator defined as $N(n)=\sum_{\alpha\sigma}a^{\dagger}_{n\alpha\sigma}a_{n\alpha\sigma}$. ${\bf S}$ and $T_{\mu}$ are the localized spin one and orbital triplet operators of $n_d=6$ space, respectively. $\mu_{\rm eff}=-\epsilon+(\tilde{U}'-\tilde{J})/2$ and $U_{\rm eff}=(\tilde{U}'-\tilde{J})/2$ are the effective chemical potential and Coulomb interaction. The hopping of quasiparticle ($a_{n\alpha\sigma}$) is renormalized to a smaller value about $\sim 400$ K at most by two factors. One is the overlap of $T_2^{(-)}$ molecular orbital with a $t_{2g}$ atomic orbital on one site, while the other is the renormalization factor of quasiparticle ($Z\sim 0.8$ for $U=1.5$ eV). Precisely speaking, exchange processes are present not only for pairs of $n_d=6$ configurations but also other configurations with different $n_d$, but we consider in the model (\ref{efftJ}) only the former ones, since they are dominant. As discussed before the exchange couplings $\tilde{J}_a^{\mu\nu}$ ($a=1$ and 2) are slightly different from $J_a^{\mu\nu}$ in Eq. (\ref{ExHamil}), but their effects are essentially the same as before. The difference is that virtual processes via the ground states of $n_d=5$ or 7 configurations are now not counted for $\tilde{J}_a^{\mu\nu}$. For example, ferromagnetic contributions in spin exchange couplings are reduced leading to $\tilde{J}^{00}_2 > J^{00}_2$. 

As we noted above, there exist competing interactions some of which 
favor magnetic ground states, while the others stabilize nonmagnetic
 states. Moreover, the magnetic interactions are strongly 
correlated with the orbital ones. There are thirty-fold degeneracies 
in the orbital configurations in the case of four coupled tetrahedra.
Due to spin-orbital couplings these degeneracies are lifted and we
 investigated which pair of orbitals favors ferro- or antiferro-magnetic
 correlations. These competitions are 
controlled particularly by the energy 
level of $A_1$ molecular orbital. This is because 
ferromagnetic spin exchange is generated by virtual hopping processes 
including $A_1$ orbital and its coupling constant is enhanced when 
$A_1$ energy level becomes lower. The four-tetrahedron calculations 
in Sec. \ref{4tetPhasediagram}
 showed that tetrahedron degrees of freedom (spin $1$ and orbital $T_1$) 
are partially screened by the exchange interactions, which leads to 
nonmagnetic $^1\!E$ ground states. 
It is quite likely that the heavy fermion behaviors of 
LiV$_2$O$_4$ stem from these competitions. Low-temperature
 metallic behaviors in LiV$_2$O$_4$ are dominated by correlated 
one-particle excitations. We expect that these competing fluctuations
 in spin and orbital also, influence the coherence of electron dynamics 
and strongly renormalize their quasiparticle weight. A part of the 
renormalization already comes from fast dynamics in the tetrahedron unit discussed in Sec. \ref{sec:Effec1tet} ($Z\sim 0.80$ $(0.66)$
 for $U=1.5$ $(3.0)$ eV  as a tetrahedron unit). It is expected that the 
quasiparticle weight $Z$ is further renormalized to a much smaller value when the effects of low-energy excitations in the  effective model (\ref{efftJ}) are fully taken into account. We expect that due to the competing interactions
 in (\ref{efftJ}), the low-temperature quasiparticles (if obtained) 
are dressed by the spin, orbital and spin-orbital interactions and thus become heavy fermions.

\section{\label{sec:Conclusion2}SUMMARY}

In the following, we review this paper as a summary.
In this paper, we have investigated the three-orbital 
Hubbard model on the pyrochlore lattice in order to 
study the heavy fermion 
behaviors of LiV$_2$O$_4$.  To study which type of 
degrees of freedom plays an important role in low-energy 
dynamics of this model, we have employed an approach of 
real-space renormalization group type. In the first stage of coarse 
graining, block variables are defined as follows for each primitive unit cell
 of pyrochlore lattice, i.e., a tetrahedron 
composed of four vanadium atoms.  

First we numerically diagonalized the three-orbital Hubbard model
 and calculated the ground state and low-energy 
excited states in this unit for the cases of electron numbers 
from $n_d$=4 to 7.  The case of $n_d$=6 corresponds 
to the average density in LiV$_2$O$_4$ ($d^{1.5}$ per 
vanadium atom), and other cases describe charge 
excitations.  One important result is that these 
low-energy states can be represented very precisely by a 
simple picture of molecular orbitals.  
The ground state of the $n_d$=4 case has a closed shell 
electron configuration of the lowest 
molecular orbital $E$.  The ground states of the $n_d$=5, 6, 
and 7 cases are described as the fully occupied $E$-orbitals 
plus partially occupied $T_2^{(-)}$-orbitals in which  
electron spins are polarized due to ferromagnetic 
Hund coupling.  

Secondly, we derived an effective Hamiltonian for 
coupled tetrahedra as for the next stage of the 
renormalization group procedure.  We have performed 
this, particularly for the case of 24 electrons in 
four coupled tetrahedra, 
which corresponds to 16 vanadium atoms constituting the cubic 
unit cell of the original pyrochlore lattice.  
This is also a natural choice of unit for block transformation 
in the second stage of the renormalization group approach, and 
we have calculated the ground state and a few 
lowest excited states of the effective Hamiltonian 
by numerical diagonalization.  
One important result is that there appear three 
types of ground states in a realistic region of 
parameters in the Hamiltonian and also that each of them 
is degenerate either in the orbital sector (${}^{1}\! E$), 
in the spin sector (${}^{7}\! A_1$) or in both sectors 
(${}^{3}\! T_1$).  It is also important that these three 
types of states are nearly degenerate to each other, and those 
that are not the ground state are the lowest and 
the second lowest excited-state multiplets.

Thirdly, we examined in detail which processes are 
important for stabilizing these low-energy states in 
the four tetrahedra.  There are two types of processes: 
one is a kinetic term and the other is interaction.  
The former is the process of electron hoppings 
from one tetrahedron to another.  
The interaction processes do not change the electron 
number in each tetrahedron but do change spin and/or orbital 
configurations.  We determined the amplitudes of effective 
electron hopping between a nearest neighbor pair of tetrahedra 
and found that they are the renormalized to a small value, 
$\sim 0.045$ eV. Since the effective hopping is small, the interactions are 
short ranged in space and the dominant ones are exchange processes of 
spin and orbital degrees of freedom between nearest neighbor 
tetrahedron pairs. In this effective exchange process, each tetrahedron 
is assumed to have six electrons and its electron 
configuration takes one of the degenerate ${}^{3}T_1$ ground states; 
i.e., three-fold orbital degrees of freedom 
and spin $S=1$ remain.  
Other tetrahedron configurations are taken into account 
only as virtual intermediate states of the exchange processes 
and they are traced out.  
The interaction Hamiltonian consists of 
pure spin exchanges, pure orbital exchanges and also 
simultaneous exchanges of spin and orbital.  
We used symmetry arguments to simplify this interaction Hamiltonian 
and determined its form. Spin space is isotoropic in our starting 
microscopic Hamiltonian and therefore the spin exchange is 
Heisenberg type.  Orbital space is not isotropic, but 
there are constraints in the orbital exchanges due to 
the symmetries of the lattice and the orbital wavefunctions 
along with the time reversal symmetry.
 As a result, the pure orbital exchanges are simplified 
to 13 independent coupling constants.  
Including the pure spin exchange and spin-orbital couplings, 
the effective exchange Hamiltonian has 34 coupling constants 
in total.  They are functions of the microscopic 
parameters and we numerically determined their values by 
carrying out the second-order perturbation 
in inter-tetrahedron hopping.  

Fourthly, we calculated the ground state and low-energy states 
of the spin-orbital exchange model, particularly for 
the unit of four tetrahedra.  We found that two sets of 
special orbital configurations are stabilized by the 
dominant term of the orbital exchange part.  They are 
further coupled to each other by subdominant 
orbital exchange processes to form three low-energy 
orbital multiplets.  These three orbital multiplets are 
also coupled with spin wavefunctions and form spin-orbital
 states in low-energy region. There, spin-orbital wavefunctions are entangled in orbital and spin spaces. This manifests strong coupling of 
spin and orbital degrees of freedom. The ground states obtained
 in this spin-orbital exchange model qualitatively agree with those
 obtained in Sec. \ref{4tetPhasediagram}. This means that the 
overall properties of this system are determined by local spin 
and orbital degrees of freedom. 

Finally, combining these results, 
we have proposed a low-energy effective model for LiV$_2$O$_4$ 
in Sec. \ref{sec:Conclusion1}. The effective model proposed contains the 
competitions of double- and super-exchange magnetic interactions coupled
 with orbital degrees of freedom. Using this effective model, we have
 discussed the origin of heavy fermion behaviors in LiV$_2$O$_4$. 
 To explain heavy fermion behaviors, it is important to identify
 the origins of large entropy at low temperatures. In our effective model, the 
entropy arises mainly from the finite spin $(S=1)$ and orbital (triplet) at each tetrahedron of the effective f.c.c. lattice. Usually (typically insulating systems with spin or orbital moments), these degrees of freedom undergo phase transitions. In our effective model, the spin or orbital moments cannot order due to the competitions of interactions. In addition to this, the geometrical frustrations in the effective f.c.c. lattice would also suppress phase transitions. This means that, after integrating out high energy incoherent excitations in the first renormalization group step, there are still a lot of low-lying incoherent spin and orbital excitations down to low temperatures and these excitations prevent quasiparticles formed. From these, it is expected that the system evolves Fermi surfaces and exhibits heavy fermion behaviors below a characteristic temperature, at which well-defined quasiparticles appear, that would be suppressed by these interactions. Interestingly, an insulating phase is found at high pressure\cite{26}. This implies that there are competing interactions in LiV$_2$O$_4$ at ambient pressure. It is an open question and interesting to explore the microscopic aspect of this transition and the relation between the heavy fermion behaviors. It is important to analyze the low-energy fluctuations in the effective model (\ref{efftJ}) to see whether a heavy fermi liquid state is realized. Elaborate large scale simulations are desired for better understanding of this model and remain as a future problem.

As an implication of the present approach, we make a comment on the temperature dependence of susceptibility. In Ref. 28, an independent tetrahedron description was applied to fit the susceptibility data at high temperatures. We can examine this point by calculating the energy change of the ground state when four tetrahedra are coupled and it is estimated to be $\simeq 400$ K per tetrahedron. This scale is not larger than the crossover temperature of the susceptibility ($T_{\rm cross}\simeq 500$ K for $J=0.2$ and $0.3$ eV) and therefore our arguments based on isolated tetrahedron remain qualitatively valid, and the crossover is mainly due to the suppression of charge fluctuations. Of course, inter-tetrahedron spin correlations also contribute to the temperature dependence of magnetic moments and this is also an important future problem.
 
We make another comment on the scenarios of the Kondo effect or the Mott transition. In these scenarios, localized $a_{1g}$ orbitals play an important role to explain the heavy fermion behaviors of LiV$_2$O$_4$. Our result is not consistent to such a situation. In the realistic parameter space, our calculations show that the density of $a_{1g}$ electron is far below unity per site in the low-energy sector. 
This feature is not consistent with these scenarios where the essential point of physics lies in the half filled configuration of $a_{1g}$ orbital. Experimentally, as observed by Jonss\"on {et al.}, LiV$_2$O$_4$ remains a bad metal at high temperature.\cite{7} Moreover there is no signature of logarithmic increase in the resistivity in the whole temperature region. These results do not support the Kondo scenario in LiV$_2$O$_4$ either.

In conclusion, we have investigated an effective Hamiltonian of three-orbital Hubbard model on a pyrochlore lattice. We have discussed the inter-tetrahedron correlations and one particle excitations by carrying out two-stage real space renormalization group calculations: a tetrahedron unit and then four coupled tetrahedra. We have concentrated on $^3T_1$ phase of one tetrahedron which has spin-one and orbital-triplet ground states. It is found that the one-particle excitations in $^3T_1$ phase are described by only $T_2^{(-)}$ molecular orbital even in the strongly correlated regime. We have derived an effective exchange model in the form of Kugel-Khomskii model with spin one and orbital triplet. Low-energy orbital correlations are analyzed together with spin-orbital correlations. It is found that orbital correlations are strongly coupled with spin correlations. Finally, we have proposed an effective Hamiltonian for LiV$_2$O$_4$ similar to a $t$-$J$ model, in which there are competing ferro- and antiferro-magnetic interactions coupled with orbital configurations together with mobile electrons. These competing interactions are expected to generate a new small energy scale and becomes an origin of heavy quasiparticles with cooperating with geometrical frustration of the pyrochlore lattice. These results would provide a good starting point for the further studies of the renormalization group analysis to understand the exotic properties in LiV$_2$O$_4$.

\begin{acknowledgments}
The authors thank S. Niitaka for sending his unpublished data. A part of the numerical computations was done at the Supercomputer Center at ISSP, University of Tokyo. This work was partly supported by KAKENHI(No. 19052003, No. 17071011 and No. 20740189) and also by the Next Generation Super Computing Project, Nanoscience Program, from the MEXT of Japan.

\end{acknowledgments}
\vspace{.5cm}
\appendix

\section{\label{sec:1particleorbital}ONE PARTICLE ORBITAL}
In this Appendix we show the wavefunctions for the one-particle molecular orbitals. There are twelve states as molecular orbitals for one tetrahedron in our model: $A_1$, $E$, $T_1$ and 2$T_2$. Since there are two kinds of $T_2$ orbitals, these two states can mix with each other. The d-electron annihilation operators in the molecular orbital basis $d_{\Gamma}$ are given as follows (we omit the site and spin indices).

\begin{widetext}
\begin{eqnarray}
   d_{A_1}&=&\frac{1}{2\sqrt{3}} \sum_{n=1}^4[ \alpha_nd_{nyz}+\beta_nd_{nzx}+\gamma_nd_{nxy}]
,\label{dA1}\\
   d_{E_{x^2-y^2}}&=&\frac{1}{2\sqrt{2}}\sum_{n=1}^4[\alpha_nd_{nyz}-\beta_nd_{nzx}
],\\
   d_{E_{3z^2-r^2}}&=&\frac{1}{2\sqrt{6}}\sum_{n=1}^4[2\gamma_nd_{nxy}-\alpha_nd_{nyz}-\beta_nd_{nzx}
],\nonumber\\
\\
\left \{
\begin{array}{@{\,}c@{\,}}
   d_{T_{1a}}\\
   d_{T_{2a}^{(2)}}
\end{array}
\right \}
&=&\frac{1}{2\sqrt{2}}\sum_{n=1}^4[ \mp\alpha_nd_{nzx}+\beta_nd_{nyz}
],\\
\left \{
\begin{array}{@{\,}c@{\,}}
   d_{T_{1b}}\\
   d_{T_{2b}^{(2)}}
\end{array}
\right \}
&=&\frac{1}{2\sqrt{2}}\sum_{n=1}^4[\mp\beta_nd_{nxy}+\gamma_nd_{nzx}
],\\
\left \{
\begin{array}{@{\,}c@{\,}}
   d_{T_{1c}}\\
   d_{T_{2c}^{(2)}}
\end{array}
\right \}
&=&\frac{1}{2\sqrt{2}}\sum_{n=1}^4[\mp\gamma_nd_{nyz}+\alpha_nd_{nxy},
],\\
\left \{
\begin{array}{@{\,}c@{\,}}
d_{T_{2a}^{(1)}}\\
d_{T_{2b}^{(1)}}\\
d_{T_{2c}^{(1)}}
\end{array}
\right \}
&=&\frac{1}{2}\sum_{n=1}^4
\left \{
\begin{array}{@{\,}c@{\,}}
d_{nxy}\\
d_{nyz}\\
d_{nzx}
\end{array}
\right \},\label{dT2c2}
\end{eqnarray}
\end{widetext}
where the signs are $(\{\alpha_n\})=(+,-,+,-)$, $(\{\beta_n\})=(+,-,-,+)$ and $(\{\gamma_n\})=(+,+,-,-)$. Note that ${\bf r}_n \equiv (\alpha_n,\beta_n,\gamma_n)$ coincides with the direction from the site $n$ to the center of the tetrahedron. We label three states of $T_1$ and two $T_2$ representations such that $(T_{1a},T_{1b},T_{1c})\propto((xy+c_1 z)(x^2-y^2),(yz+c_1 x)(y^2-z^2),(zx+c_1 y)(z^2-y^2))$, and $(T^{(n)}_{2a},T^{(n)}_{2b},T^{(n)}_{2c})\propto((xy+c_2^{(n)} z),(yz+c_2^{(n)} x),(zx+c_2^{(n)} y))$ for $n=1\ {\rm and}\ 2$, where $c_1$ and $c_2^{(n)}$ are constants. The site indices on the right hand side of Eqs. (\ref{dA1})-(\ref{dT2c2}) are those in a unit cell and indicated in Fig. {\ref{fig-lat}}(a).

\section{ORBITAL WAVEFUNCTIONS}\label{AppOrbWavefunc}

\begin{figure}[tb]
  \begin{center}
    \includegraphics[width=.4\textwidth]{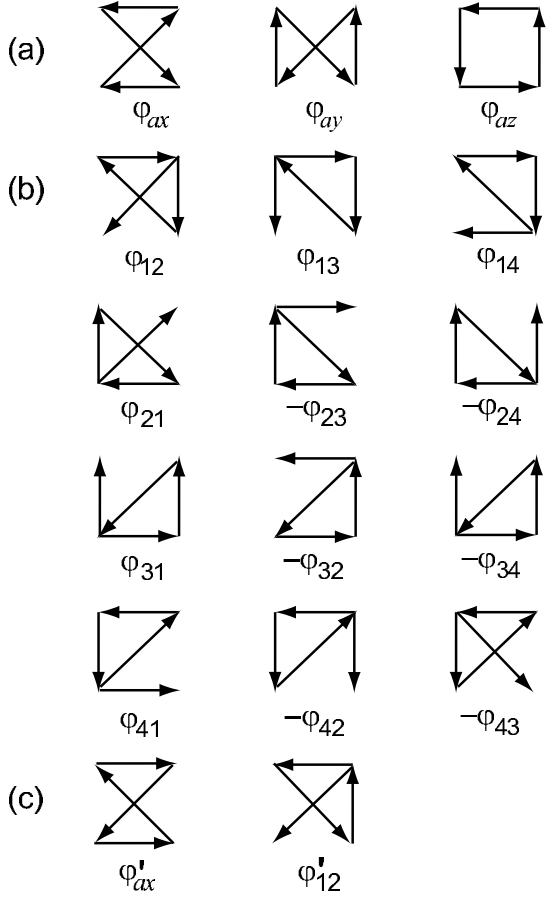}
  \end{center}
\caption{Orbital basis wavefunctions. (a) closed path graphs. (b) lasso graphs. See also Fig. \ref{fig-closetriangle}. (c) examples of $\varphi'=\hat{R}\varphi$.}
\label{fig-orbgraph1}
\end{figure}
%

In this Appendix, we study the orbital part 
of the exchange model (\ref{FullEX}) and explain in detail 
the calculation of low-energy eigenstates 
in the unit of four coupled tetrahedra.  
As discussed in Sec. \ref{sec:orbwave}, two sets of 
states are favored by the largest term $J_1^{88}$ 
of the orbital couplings: 6 states of type (a) 
and 24 states of type (b).  Half of them are 
shown in Fig. \ref{fig-orbgraph1} (a) and (b) with the 
arrow representation explained in Fig. \ref{fig-closetriangle}. 
Some of these are defined with minus sign 
as a phase factor for later convenience.  
The other half of the states are defined 
by reversing the direction of arrows in 
the part of closed path and denoted with 
prime symbol like $\varphi'$ as shown in Fig. \ref{fig-orbgraph1} (c).  
We solve the eigenvalue problem of the orbital 
exchange Hamiltonian in the subspace of 
these thirty states of type (a) and (b).

The cluster of the coupled four tetrahedra has 
also a tetrahedral symmetry $T_d$ and this 
is useful to simplify the eigenvalue problem.  
As explained in Sec. \ref{sec:orbwave}, the six states 
of type (a) are classified to three 
irreducible representations, $A_1 \oplus E \oplus T_1$ 
and they are given as 
\begin{eqnarray}
&&|a,A_1 \rangle = {\textstyle \frac{1}{\sqrt{6}}}
(1+ \hat{R}) (\varphi_{ax}+\varphi_{ay}+\varphi_{az}) , 
\\
&&|a,E_{x^2-y^2} \rangle = {\textstyle \frac{1}{2}}
(1+ \hat{R}) (\varphi_{ax}-\varphi_{ay}) , 
\\
&&|a,T_{1a} \rangle = {\textstyle \frac{1}{\sqrt{2}}}
(1- \hat{R}) \varphi_{ax} .  
\end{eqnarray}
Here $\hat{R}$ is the operator that reverses 
the arrow direction in the closed path part, 
i.e., $\hat{R} \varphi = \varphi'$, and 
the other basis states of the $E$- and $T_1$-representations 
are obtained by applying appropriate symmetry operations 
to these.  Similarly, 24 states of type (b) are 
classified to 
$A_1 \oplus A_2 \oplus 2E \oplus 3T_1 \oplus 3T_2$ 
and the representatives of their basis states are 
\begin{eqnarray}
&&\left.\begin{array}{c}
|b,A_1 \rangle \\ 
|b,A_2 \rangle \\ 
\end{array} 
\right\}
= {\textstyle \frac{1}{\sqrt{24}}}
(1 \pm \hat{R}) \sum_{i \ne j}\varphi_{ij} , 
\\
&&\left.\begin{array}{c} 
|b,E_{x^2-y^2}^{(1)} \rangle \\ 
|b,E_{x^2-y^2}^{(2)} \rangle 
\end{array}
\right\}
= {\textstyle \frac{1}{4}}
(\hat{R} \pm 1) \left(
\varphi_{14}-\varphi_{13} 
+\varphi_{23}-\varphi_{24}
\right. 
\nonumber\\
&&\left. \hspace{2.5cm}
+\varphi_{32}-\varphi_{31}
+\varphi_{41}-\varphi_{42}
\right) , 
\\
&&\left.\begin{array}{c} 
|b,T_{1a}^{(1)} \rangle \\ 
|b,T_{2a}^{(1)} \rangle 
\end{array}
\right\}
= {\textstyle \frac{1}{\sqrt{8}}}
(1 \mp \hat{R}) \left(
\varphi_{13}
-\varphi_{24}
+\varphi_{31}
-\varphi_{42}\right) , 
\\
&&\left.\begin{array}{c} 
|b,T_{1a}^{(2)} \rangle \\ 
|b,T_{2a}^{(2)} \rangle 
\end{array}
\right\}
= {\textstyle \frac{1}{\sqrt{8}}}
\left[
\left(
\varphi_{14}
-\varphi_{23}
+\varphi_{32}
-\varphi_{41} \right)
\right. 
\nonumber\\
&&\left. \hspace{2.5cm}
\mp \hat{R} 
\left(
\varphi_{12}
-\varphi_{21}
+\varphi_{34}
-\varphi_{43} \right)
\right] , 
\\
&&\left.\begin{array}{c} 
|b,T_{1a}^{(3)} \rangle \\ 
|b,T_{2a}^{(3)} \rangle 
\end{array}
\right\}
= {\textstyle \frac{1}{\sqrt{8}}}
\left[
\left(
\varphi_{12}
-\varphi_{21}
+\varphi_{34}
-\varphi_{43}\right)
\right. 
\nonumber\\
&&\left. \hspace{2.5cm}
\mp \hat{R}
\left(
\varphi_{14}
-\varphi_{23}
+\varphi_{32}
-\varphi_{41}\right)
\right] . 
\end{eqnarray}
The other basis states are also generated by 
applying appropriate symmetry operations. 

The orbital exchange Hamiltonian for the four tetrahedra 
\begin{equation} 
H_{\rm orb} = \sum_{1 \le n<m \le 4} \sum_{\mu,\nu=1}^8 
{\textstyle \frac{2}{3}} J_1^{\mu\nu}(mn) 
T_{\mu}(m) T_{\nu}(n)
\end{equation}
has finite 
matrix elements only between the basis states 
in the same representation. In the subspace of thirty states 
of type (a) and (b), some pairs of coupling constants are not 
independent and it is convenient to introduce the parameters 
$K_{\pm}\equiv(J_1^{11}+J_1^{55})\pm(J_1^{22}+J_1^{66})$. 
The $J_1^{88}$ term gives a constant energy $E_0=-6J_1^{88}$ 
in this subspace. Aside from this constant, the results 
are the following. 

\paragraph{$A_1$-representation: (dimension 2)}
\begin{widetext}
\begin{eqnarray}
&&\langle a , A_1 | H_{\rm orb} | a, A_1 \rangle = 
{\textstyle \frac43} J_1^{33},
\\
&&\langle b , A_1 | H_{\rm orb} | b, A_1 \rangle = 
- K_+ - 3 \beta J_1^{18} + \alpha J_1^{17} 
 - \alpha J_1^{36}
- \alpha J_1^{42},
\\
&&\langle a , A_1 | H_{\rm orb} | b, A_1 \rangle = 
{\textstyle \frac23} K_- - 2\gamma J_1^{78} + 2\beta J_1^{18} 
- \alpha J_1^{17}  - \alpha J_1^{42},
\end{eqnarray}
where $\alpha=\sqrt{8}/3,  \beta = (2/3 )^{3/2}$ and $\gamma=4/\sqrt{27}$. 

\paragraph{$A_2$-representation: (dimension 1)}
\begin{eqnarray}
&&\langle b , A_2 | H_{\rm orb} | b, A_2 \rangle = 
{\textstyle \frac13} K_+ - 2\gamma J_1^{78} - \beta J_1^{18} -  \alpha J_1^{17}  + \alpha J_1^{36} + \alpha J_1^{42}.
\end{eqnarray}

\paragraph{$E$-representation: (dimension 3)}
\begin{eqnarray}
&&\langle a , E | H_{\rm orb} | a, E \rangle = 
{\textstyle \frac43} J_1^{33},
\\
&&\langle b , E^{(1)} | H_{\rm orb} | b, E^{(1)} \rangle = 
- K_+ + {\textstyle \frac32} \beta J_1^{18} - {\textstyle \frac12} \alpha J_1^{17}
+ {\textstyle \frac12} \alpha J_1^{36} + {\textstyle \frac12} \alpha J_1^{42},
\\
&&\langle b , E^{(2)} | H_{\rm orb} | b, E^{(2)} \rangle =
{\textstyle \frac13} K_+ + \gamma J_1^{78} + {\textstyle \frac12} \beta J_1^{18} 
+ {\textstyle \frac12} \alpha J_1^{17}
- {\textstyle \frac12} \alpha J_1^{36} 
- {\textstyle \frac12} \alpha J_1^{42}, 
\\
&&\langle a , E | H_{\rm orb} | b, E^{(1)} \rangle = 
{\textstyle \frac13} K_- - \gamma J_1^{78} + \beta J_1^{18} + \alpha J_1^{17} 
 - {\textstyle \frac16} J_1^{33} + \alpha J_1^{42},
\\
&&\langle a , E | H_{\rm orb} | b, E^{(2)} \rangle = 
{\textstyle \frac{1}{\sqrt{3}}} (
K_- + 3 \gamma J_1^{78} -3 \beta J_1^{18}),
\\
&&\langle b , E^{(1)} | H_{\rm orb} | b, E^{(2)} \rangle = 
{\textstyle \frac{\sqrt{3}}{2}} (\gamma J_1^{78}-\beta J_1^{18} 
-\alpha J_1^{17} -\alpha J_1^{36}+\alpha J_1^{42}).
\end{eqnarray}
\end{widetext}
\paragraph{$T_1$-representation: (dimension 4)}
\begin{eqnarray}
&&\langle a , T_1 | H_{\rm orb} | a, T_1 \rangle = 
{\textstyle \frac43} J_1^{33}, 
\\
&&\langle b , T_1^{(1)} | H_{\rm orb} | b, T_1^{(1)} \rangle = 
-{\textstyle \frac13} K_+,
\\
&&\langle b , T_1^{(2)} | H_{\rm orb} | b, T_1^{(2)} \rangle = 
({\textstyle \frac13} K_+ -  \gamma J_1^{78} +  \beta J_1^{18} 
+  \alpha J_1^{36} ),\nonumber\\
\\
&&\langle b , T_1^{(3)} | H_{\rm orb} | b, T_1^{(3)} \rangle = 
-{\textstyle \frac13} (
K_+ -3 \alpha J_1^{17} +3  \alpha J_1^{42}
),
\\
&&\langle a , T_1 | H_{\rm orb} | b, T_1^{(1)} \rangle = 
 - \alpha ( J_1^{17} + J_1^{42} ),
\\
&&\langle a , T_1 | H_{\rm orb} | b, T_1^{(2)} \rangle = 
-{\textstyle \frac23} K_-,
\\
&&\langle a , T_1 | H_{\rm orb} | b, T_1^{(3)} \rangle = 
2 \gamma J_1^{78} - 2 \beta J_1^{18},
\\
&&\langle b , T_1^{(1)} | H_{\rm orb} | b, T_1^{(2)} \rangle = 
{\textstyle \frac12} \gamma J_1^{78} + \beta J_1^{18} ,
\\
&&\langle b , T_1^{(1)} | H_{\rm orb} | b, T_1^{(3)} \rangle = 
- {\textstyle \frac12} \gamma J_1^{78} + 2 \beta J_1^{18} + \alpha J_1^{36},
\\
&&\langle b , T_1^{(2)} | H_{\rm orb} | b, T_1^{(3)} \rangle = 
- {\textstyle \frac12} \gamma J_1^{78} - \beta J_1^{18}.
\end{eqnarray}

\paragraph{$T_2$-representation: (dimension 3)}
\begin{eqnarray}
&&\langle b , T_2^{(1)} | H_{\rm orb} | b, T_2^{(1)} \rangle = 
K_+, 
\\
&&\langle b , T_2^{(2)} | H_{\rm orb} | b, T_2^{(2)} \rangle = 
{\textstyle \frac13} K_+ + \gamma J_1^{78} 
- \beta J_1^{18}  - \alpha J_1^{36}, \nonumber\\
\\
&&\langle b , T_2^{(3)} | H_{\rm orb} | b, T_2^{(3)} \rangle = 
- {\textstyle \frac13} K_+ - \alpha J_1^{17} + \alpha J_1^{42},
\\
&&\langle b , T_2^{(1)} | H_{\rm orb} | b, T_2^{(2)} \rangle = 
{\textstyle \frac12} \gamma J_1^{78} + \beta J_1^{18},
\\
&&\langle b , T_2^{(1)} | H_{\rm orb} | b, T_2^{(3)} \rangle = 
{\textstyle \frac32} \gamma J_1^{78} - \alpha J_1^{36},  
\\
&&\langle b , T_2^{(2)} | H_{\rm orb} | b, T_2^{(3)} \rangle = 
- {\textstyle \frac12} \gamma J_1^{78} - \beta J_1^{18}. 
\end{eqnarray}

Thus the Hamiltonian is reduced to small matrices and 
the largest size of matrix is four.  It is possible to 
obtain analytic expressions of the eigenenergies, but 
we do not write here very lengthy results.

\section{WAVEFUNCTIONS FOR FOUR S=1 SPINS ON A TETRAHEDRON}\label{AppSpinWavefunc}
\begin{figure}[t!]
  \begin{center}
    \includegraphics[width=.45\textwidth]{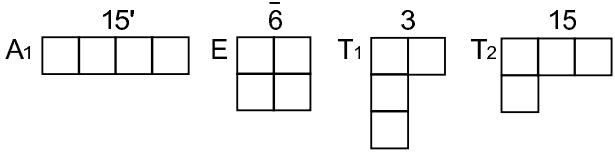}
  \end{center}
\caption{Young diagrams for symmetric group $S_4$.}
\label{fig-youngdiagram}
\end{figure}

 In this appendix, we show the spin wavefunctions on a tetrahedron constructed of four spin $S=1$. These wavefunctions are classified by the total spin $S$ and the irreducible representation $\Gamma$ of $T_d$ point group and listed in Table \ref{tbl-Spintable}. The point group $T_d$ is isomorphic to the symmetric group $S_4$ when permutations of tetrahedron vertices are concerned, and therefore Young diagrams can alternatively be used for irreducible representations, see Fig. \ref{fig-youngdiagram}. This is useful particularly when we see the symmetries of the wavefunctions. 

In the following, the spin wavefunctions are represented by the linear combination of $|s_z(1)s_z(2)s_z(3)s_z(4)\rangle$, where $s_z(n)$$(=-1, 0$ and $1)$ represents the eigenvalue for the z-component of the spin at the site (tetrahedron) $n$ in Fig. \ref{fig-lat} (b). For convenience, we write $-1$ as $\bar{1}$ and list the highest states ($S_z=S$) below.

\begin{table}[!t]
\caption{List of the spin wavefunctions on a tetrahedron constructed by four spin-1 states.}
\begin{ruledtabular}
  \begin{tabular}{rl}
     $S$ & $\Gamma$ \\
\hline
     $4$   & $A_1$(singlet) \\
     $3$   & $T_2$(triplet) \\
     $2$   & $A_1$(singlet), $E$(doublet), $T_2$(triplet) \\
     $1$   & $T_1$(triplet), $T_2$(triplet) \\
     $0$   & $A_1$(singlet), $E$(doublet)
        \end{tabular}
\end{ruledtabular}
\label{tbl-Spintable}
\end{table}
%

\subsection{Wavefunctions for $S=3$ and $4$}\label{AppS=3}
Although we do not discuss the spin wavefunction with $S=3$ and $4$ in this paper, we list the form of the wavefunctions for completeness.

$S=4$, $S_z=4$
\begin{eqnarray}
|^9\!A_1\rangle&=& |1111\rangle.  
\end{eqnarray}

$S=3$, $S_z=3$
\begin{eqnarray}
|^7T_{2a}\rangle&=&\!\! \frac{1}{2}\Big[ |1110\rangle +|1101\rangle -|1011\rangle -|0111\rangle \Big],\\
|^7T_{2b}\rangle&=&\!\! \frac{1}{2}\Big[ |1110\rangle -|1101\rangle +|1011\rangle -|0111\rangle \Big],\\
|^7T_{2c}\rangle&=&\!\! \frac{1}{2}\Big[ -|1110\rangle +|1101\rangle +|1011\rangle -|0111\rangle \Big] .
\end{eqnarray}

\subsection{Wavefunctions for $S=2$, $S_z=2$}\label{AppS=2}
The spin wavefunctions for $S=2$ are used for discussing the $^5\!A_2$ spin-orbital ground states of four tetrahedra in Sec. \ref{(0E)}.
\begin{eqnarray}
|^5A_{1}\rangle&=& \frac{\sqrt{6}}{2\sqrt{7}}\Big[ |111\bar{1}\rangle + {\rm permutations} \Big]\nonumber\\
&&-\sqrt{\frac{1}{21}}\Big[ |0011\rangle +\ {\rm permutations}\Big],\\
|^5E_{x^2-y^2}\rangle&=&\frac{1}{2}\Big[ |1010\rangle -|1001\rangle -|0110\rangle +|0101\rangle \Big],\\
|^5E_{3z^2-r^2}\rangle&=&\frac{1}{2\sqrt{2}}\Big[ 2|1100\rangle +2|0011\rangle-|1001\rangle -|1010\rangle\nonumber\\
&&-|0110\rangle -|0101\rangle \Big],\\
|^5T_{2a}\rangle&=& \frac{1}{\sqrt{6}}\Big[ |111\bar{1}\rangle +|11\bar{1}1\rangle -|1\bar{1}11\rangle -|\bar{1}111\rangle \Big]\nonumber\\
&&- \frac{1}{\sqrt{6}}\Big[ |1100\rangle -|0011\rangle \Big],\\
|^5T_{2b}\rangle&=& \frac{1}{\sqrt{6}}\Big[ |111\bar{1}\rangle -|11\bar{1}1\rangle +|1\bar{1}11\rangle -|\bar{1}111\rangle \Big]\nonumber\\
&&- \frac{1}{\sqrt{6}}\Big[ |1010\rangle -|0101\rangle \Big],\\
|^5T_{2c}\rangle&=& \frac{1}{\sqrt{6}}\Big[ -|111\bar{1}\rangle +|11\bar{1}1\rangle +|1\bar{1}11\rangle -|\bar{1}111\rangle \Big]\nonumber\\
&&- \frac{1}{\sqrt{6}}\Big[ |1001\rangle -|0110\rangle \Big].
\end{eqnarray}

\subsection{Wavefunctions for $S=1$, $S_z=1$}\label{AppS=1}
The spin wavefunctions for $S=1$ are used for discussing the $^3T_1$ spin-orbital ground states of four tetrahedra in Sec. \ref{(0E)}. The wavefunctions are given as
\begin{widetext}
\begin{eqnarray}
|^3T_{1a}\rangle&=& \frac{1}{2\sqrt{2}}\Big[ |0\bar{1}11\rangle -|\bar{1}011\rangle +|110\bar{1}\rangle -|11\bar{1}0\rangle +|\bar{1}110\rangle -|011\bar{1}\rangle +|10\bar{1}1\rangle -|1\bar{1}01\rangle\Big],\\
|^3T_{1b}\rangle&=& \frac{1}{2\sqrt{2}}\Big[ |01\bar{1}1\rangle -|\bar{1}101\rangle +|1\bar{1}10\rangle -|101\bar{1}\rangle +|\bar{1}011\rangle -|0\bar{1}11\rangle +|110\bar{1}\rangle -|11\bar{1}0\rangle\Big],\\
|^3T_{1c}\rangle&=& \frac{1}{2\sqrt{2}}\Big[ |011\bar{1}\rangle -|\bar{1}110\rangle +|10\bar{1}1\rangle -|1\bar{1}01\rangle + |\bar{1}101\rangle -|01\bar{1}1\rangle +|1\bar{1}10\rangle -|101\bar{1}\rangle \Big],\\
|^3T_{2a}\rangle&=& \frac{1}{\sqrt{10}}\Big[ |0001\rangle +|0010\rangle -|0100\rangle -|1000\rangle \Big] -\frac{1}{\sqrt{10}}\Big[ |0\bar{1}11\rangle +|\bar{1}011\rangle -|110\bar{1}1\rangle -|11\bar{1}0\rangle\Big]\nonumber\\
&&-\frac{1}{2\sqrt{10}}\Big[ |011\bar{1}\rangle -|\bar{1}110\rangle +|10\bar{1}1\rangle -|1\bar{1}01\rangle+ |01\bar{1}1\rangle -|\bar{1}101\rangle +|101\bar{1}\rangle -|1\bar{1}10\rangle\Big],\\
|^3T_{2b}\rangle&=& \frac{1}{\sqrt{10}}\Big[ |0100\rangle -|0001\rangle +|0010\rangle -|1000\rangle \Big] -\frac{1}{\sqrt{10}}\Big[ |01\bar{1}1\rangle +|\bar{1}101\rangle -|1\bar{1}10\rangle -|101\bar{1}\rangle\Big]\nonumber\\
&&-\frac{1}{2\sqrt{10}}\Big[ |0\bar{1}11\rangle -|\bar{1}011\rangle +|110\bar{1}\rangle -|11\bar{1}0\rangle+ |011\bar{1}\rangle -|\bar{1}110\rangle +|1\bar{1}01\rangle -|10\bar{1}1\rangle\Big],\\
|^3T_{2c}\rangle&=& \frac{1}{\sqrt{10}}\Big[ -|0001\rangle +|0010\rangle +|0100\rangle -|1000\rangle \Big] -\frac{1}{\sqrt{10}}\Big[ |011\bar{1}\rangle +|\bar{1}110\rangle -|10\bar{1}1\rangle -|1\bar{1}01\rangle\Big]\nonumber\\
&&-\frac{1}{2\sqrt{10}}\Big[ |01\bar{1}1\rangle -|\bar{1}101\rangle +|1\bar{1}10\rangle -|101\bar{1}\rangle+ |0\bar{1}11\rangle -|\bar{1}011\rangle +|11\bar{1}0\rangle -|110\bar{1}\rangle\Big].
\end{eqnarray}

\subsection{Wavefunctions for $S=0$}\label{AppS=0}
The spin wavefunctions for $S=0$ are used for discussing the $^1\!E$ spin-orbital ground states of four tetrahedra in Sec. \ref{(0E)}. The wavefunctions are given as
\begin{eqnarray}
|^1\!A_1\rangle&=&
\frac{2}{3\sqrt{5}}\Big[ |11\bar{1}\bar{1}\rangle+ {\rm permutations} \Big]+\frac{1}{\sqrt{5}}|0000\rangle
-\frac{1}{3\sqrt{5}}\Big[ |100\bar{1}\rangle+ {\rm permutations} \Big],\\
|^1\!E_{x^2-y^2}\rangle&=&
\frac{1}{2\sqrt{3}}\sum_{s=\pm 1}\Big[ |s00\bar{s}\rangle+|0s\bar{s}0\rangle+ |s\bar{s}\bar{s}s\rangle-|s\bar{s}s\bar{s}\rangle
-|0s0\bar{s}\rangle-|s0\bar{s}0\rangle\Big],\\
|^1\!E_{3z^2-r^2}\rangle&=&
\frac{1}{6}\sum_{s=\pm 1}\Big[ 2|ss\bar{s}\bar{s}\rangle- |s\bar{s}s\bar{s}\rangle-|s\bar{s}\bar{s}s\rangle 
+2|s\bar{s}00\rangle+2|00s\bar{s}\rangle -|s00\bar{s}\rangle-|0s\bar{s}0\rangle -|0s0\bar{s}\rangle-|s0\bar{s}0\rangle\Big].
\end{eqnarray}
\end{widetext}
When we rewrite these wavefunctions by using direct products of two bond spins, {e.g.,} $S_{12}^{\mu}\equiv S_{\mu}(1)+S_{\mu}(2)$ and $S_{34}^{\mu}\equiv S_{\mu}(3)+S_{\mu}(4)$, we obtain
\begin{eqnarray}
|^1\!A_1\rangle&=&
\frac{2}{3}|22\rangle+\frac{\sqrt{5}}{3}|00\rangle,\\
|^1\!E_{x^2-y^2}\rangle&=&|11\rangle,\\
|^1\!E_{3z^2-r^2}\rangle&=&
\frac{\sqrt{5}}{3}|22\rangle-\frac{2}{3}|00\rangle.
\end{eqnarray}
Here $|S_{12}S_{34}\rangle$ represents the spin singlet state constructed from the bond state with the total spin $S_{12}$ and $S_{34}$.


\begin{thebibliography}
\bibitem{delta} Although we can fit the band structure, for example, at the $\Gamma$ point with $\Delta\sim0.1\sim0.2$ eV, the fitting results are not so different from those obtained by the parameters we use in this paper.
\bibitem{explaneplane} We choose ``a plane including the bond ($n$-$m$)'' out of $xy$, $yz$ and $zx$-planes. For example, in the case of ($1$-$3$) bond, this plane corresponds to the $yz$-plane as illustrated in Fig. \ref{fig-help} (b).
\bibitem{imag} Since the time reversal symmetry is not broken in the situation considered, coupling constants should be real. Note that $T_2$, $T_4$ and $T_5$ are pure imaginary and therefore couplings including only one of these cannot appear.
\bibitem{reduction} After the summation $n$ and $m$, terms including $T_0$ in $J_1^{\mu\nu}$ such as $T_0(n)T_8(m)$, disappear in total. There are three such terms and, neglecting a trivial constant $J_1^{00}$, the number of independent coupling constants for $J_1^{\mu\nu}$ is reduced to 13.
\end{thebibliography}

\end{document}